\documentclass[fleqn,usenatbib]{mnras}
\usepackage{adjustbox}
\usepackage{widetext}

\usepackage{adjustbox}
\usepackage{hyperref}

\def\setsymbol#1#2{\expandafter\def\csname #1\endcsname{#2}}
\def\getsymbol#1{\csname #1\endcsname}
\usepackage{amsmath}
\usepackage{amssymb}

\newbox\tablebox    \newdimen\tablewidth
\def\leaderfil{\leaders\hbox to 5pt{\hss.\hss}\hfil}

\def\tablenote#1 #2\par{\begingroup \parindent=0.8em
    \abovedisplayshortskip=0pt\belowdisplayshortskip=0pt
    \noindent
    $$\hss\vbox{\hsize\tablewidth \hangindent=\parindent \hangafter=1 \noindent
    \hbox to \parindent{$^#1$\hss}\strut#2\strut\par}\hss$$
    \endgroup}
\def\doubleline{\vskip 3pt\hrule \vskip 1.5pt \hrule \vskip 5pt}

\usepackage{newtxtext,newtxmath}

\usepackage[T1]{fontenc}

\DeclareRobustCommand{\VAN}[3]{#2}
\let\VANthebibliography\thebibliography
\def\thebibliography{\DeclareRobustCommand{\VAN}[3]{##3}\VANthebibliography}

\usepackage{graphicx}	
\usepackage{amsmath}	

\title[On the impact of $f(Q)$ gravity on the Large Scale Structure]{On the impact of $f(Q)$ gravity on the Large Scale Structure}

\author[O. Sokoliuk et al.]{
Oleksii Sokoliuk$^{1}$\thanks{oleksii.sokoliuk@mao.kiev.ua},
 Simran Arora$^{2\thanks{dawrasimran27@gmail.com }}$,
 Subhrat Praharaj$^{2\thanks{sp7437473@gmail.com}}$,
 Alexander Baransky$^{3\thanks{abaransky@ukr.net}}$,
and P.K. Sahoo$^{2\thanks{pksahoo@hyderabad.bits-pilani.ac.in}}$
\\
$^{1}$Astronomical Observatory of the National Academy of Sciences of Ukraine (MAO NASU), Kyiv, 03143, Ukraine\\
$^{2}$Department of Mathematics, Birla Institute of Technology and
Science-Pilani, Hyderabad Campus, Hyderabad-500078, India\\
$^{3}$Astronomical Observatory, Taras Shevchenko National University of Kyiv, 3 Observatorna St., 04053 Kyiv, Ukraine
}

\date{Accepted XXX. Received YYY; in original form ZZZ}

\pubyear{2015}

\begin{document}
\label{firstpage}
\pagerange{\pageref{firstpage}--\pageref{lastpage}}
\maketitle

\begin{abstract}
We investigate the exponential $f(Q)$ symmetric teleparallel gravitation, namely $f(Q)=Q+\alpha Q_0(1-e^{-\beta\sqrt{Q/Q_0}})$ using \texttt{ME-GADGET} code to probe the structure formation with box sizes $L_{\mathrm{box}}=10/100$ Mpc$/h$ and middle resolution $N_p^{1/3}=512$. To reproduce viable cosmology within the aforementioned modified gravity theory, we first perform Markov Chain Monte Carlo (MCMC) sampling on OHD/BAO/Pantheon datasets and constrain a parameter space. Furthermore, we also derive theoretical values for deceleration parameter $q(z)$, statefinder pair $\{r,s\}$ and effective gravitational constant $G_{\mathrm{eff}}$, perform $Om(z)$ diagnostics. While carrying out N-body+SPH simulations, we derive CDM+baryons over density/temperature/mean molecular weight fields, matter power spectrum (both 2/3D, with/without redshift space distortions), bispectrum, two-point correlation function and halo mass function. Results for small and big simulation box sizes are therefore properly compared, halo mass function is related to the Seth-Tormen theoretical prediction and matter power spectrum to the standard \texttt{CAMB} output.
\end{abstract}



\section{Introduction}\label{sec:1}
Numerous independent cosmological observable predicts that universe undergoes the accelerated expansion phase at the present time \citep{Riess_1998,Perlmutter_1999,Suzuki_2012,Hinshaw_2013}. It is well-known that General Theory of Relativity (GR) is a quite successful theory on various cosmological scales, and it is able to describe the recent accelerated expansion of the universe by introducing the so-called cosmological constant (or $\Lambda$ term) in the Einstein-Hilbert action integral. But, such term gives a rise to various issues, that were found in the papers of \citep{doi:10.1142/S0218271800000542,PADMANABHAN2003235}. Since the gravitational Lagrangian is not practically restricted by only the linear Ricci scalar term, one can introduce additional terms to emulate effective dark energy and reproduce different universe evolutionary phases, such as cosmological inflation or late time accelerated expansion in order to overcome the aforementioned problems.

There are a lot of different ways to modify general relativity - for example, by introducing some matter fields (canonical scalar field, vector and gauge boson fields, Dirac spinors, etc.). Another way is to present an entirely different notion of Lorentzian 4-manifold curvature by adjusting the metric-affine connection \citep{de2010f,capozziello2011extended}. For example, one could use the so-called torsion or non-metricity, which are constructed based on Weitzenb\"ock and metric incompatible affine connections, respectively. Consequently, there are two analoges to GR, namely the Teleparallel Equivalent of GR (TEGR), introduced in \citep{hayashi1979new,abedi2017nonminimal} and Symmetric Teleparallel Equivalent of GR (STEGR) \citep{Nester:1998mp,hohmann2021general}. In the current work, we will focus on the arbitrary parameterization of STEGR ($f(Q)$ gravitation) in particular.

A key aspect of $f(Q)$ theory is the usage of a flat connection pertaining to the existence of affine coordinates in which all of its components vanish, converting covariant derivatives into partial derivatives \citep{hohmann2021general,dimakis2022flrw,zhao2022covariant}. So, it is possible to distinguish gravity from inertial effects in $f(Q)$ theory. For many modified gravity theories, the development of the $f(Q)$ theory provides a fresh starting point. Additionally, it offers a straightforward formulation in which self-accelerating solutions spontaneously appear in both the early and late universe. When compared to other geometric extensions of GR, both $f(T)$ and $f(Q)$ theories have a substantial benefit in that the background field equations are always of second order, which means that Ostrogradsky's theorem \citep{motohashi2015third} related instability issues are avoided.

Up to this moment, $f(Q)$ gravity has been incorporated in dozens of studies and is a very promising theory, that can reproduce the behaviour of both early \citep{De:2022vfc} and late universe, satisfy constraints from Cosmic Microwave Background (CMB), SuperNovae (SN) distance modulus, Baryon Acoustic Oscillations (BAO), Observational Hubble Dataset (OHD) and primordial scalar index $n_s$, standard sirens from LIGO/VIRGO/ET \citep{DAgostino:2022tdk,Ferreira:2022jcd}. For instance, exponential gravitation were constrained in the study by \citep{ANAGNOSTOPOULOS2021136634,atayde2021can}, and the authors found out that such a theory can challenge concordance $\Lambda$CDM theory. There was also some studies carried out on the matter of $f(Q)$ cosmography \citep{Mandal2020} and energy conditions \citep{PhysRevD.102.024057,Koussour:2022ycn}. Additionally, observational constraints on the $f(Q)$ gravity have been established for a number of parameterizations of the $f(Q)$ function using various observational probes \citep{Lazkoz:2019sjl,Solanki:2022avt,jimenez2020cosmology}. In the context of f(Q) cosmology, a Hamiltonian formulation has been designed to carry out a canonical quantization procedure \citep{dimakis2021quantum}. Aside from these findings, $f(Q)$ gravity has been the focus of several investigations in varied applications \citep{hu2022adm,albuquerque2022designer,wang2022static,esposito2022reconstructing,bajardi2020bouncing,arora2022crossing,harko2018coupling}. In the current study, we are going to investigate the logarithmic $f(Q)$ gravity in terms of observational constraints using Markov Chain Monte Carlo (MCMC) methodologies and high resolution N-body simulations, which will be discussed in the following subsections.

\subsection{N-body simulation as a probe of modified gravity}

To probe the validity of a particular modified theory of gravitation, one needs to incorporate various cosmological observables, ranging from cosmic expansion rate to clustering and structure formation history. The latter could be most effectively studied with the use of the so-called N-body simulations, that are well-known to be the best theoretical probe of the large scale structure of the universe, that provide information on the matter power spectrum/bispectrum, $N$-point correlation functions and halo mass function, void size function etc. Over the last few years, such approach has attracted some interest in the field of modified gravity (see the work of \citep{2020MNRAS.497.1885H}). Authors of the paper \citep{Wilson:2022ets} developed a pipeline to differentiate modified gravity theories from fiducial $\Lambda$CDM model and constrain those theories properly using voids in the N-body simulations, that are known to be less affected by the non-linear and baryonic physics in relation to the dark matter halos. Besides, in addition to voids, intrinsic shape alignments of massive halos and galaxy/halo angular clustering could be used to discriminate MOG theories from $\Lambda$CDM in the presence of massive neutrino (see \citep{Lee:2022lbu} and \citep{Drozda:2022ata} respectively). Aforementioned Halo Mass Function (further-HMF) were examined for $f(Q)$ and Dvali-Gabadadze-Porrati (DGP) gravities in \citep{Gupta:2021pdy}. Widely used code \texttt{MG-Gadget}, introduced and developed in \citep{10.1093/mnras/stt1575} were employed to study $f(R)$ Hu-Sawicki theory \citep{Arnold:2016arp,Arnold:2014qha,Giocoli:2018gqh}, conformally coupled gravity \citep{Ruan:2021wup}. In turn, we are going to use \texttt{ME-GADGET} code (for documentation, check \citep{Zhang:2018glx}) to study $f(Q)$ gravity behaviour. Such code was applied to the case of $f(T)$ teleparallel theory \citep{Huang:2022slc}, interacting dark energy \citep{Zhang:2018glx,Zhao:2022ycr,Liu:2022hpz} and cubic vector gallileon \citep{Su:2022yoj}.

Our paper is organised as follows: in the first Section (\ref{sec:1}) we provide a little introduction into the topic of modified theories of gravity and N-body simulations. Consequently, in the Section (\ref{sec:2}) we present the foundations of symmetric teleparallel gravity an it's arbitrary parameterization, in the third section we adopt FLRW isotropic line element, derive field equations for our logarithmic choice of $f(Q)$ function. In the Section (\ref{sec:4}) we therefore introduce each observational dataset of our consideration and perform MCMC analysis, in Section (\ref{sec:5}) we analyze the provided constraints deriving theoretical predictions for deceleration parameter, statefinder pair and $Om(z)$. In the following section we set up the \texttt{ME-GADGET} suite and study the N-body output for small simulation box size, in (\ref{sec:7}) we therefore compare aforementioned results with the ones, obtained for large $L_{\mathrm{box}}$. Finally, in the last section we present the concluding remarks on the key topics of our study.

\section{Modified symmetric teleparallel gravitation}\label{sec:2}

Firstly we are going to start by introducing the fundamentals of the symmetric teleparallel theories of gravitation. In such theories, it is generally assumed that the scalar curvature of the manifold does vanishes (and therefore, $R=0$) as well as torsion, however non-metricity is non-zero (and describes gravitational interactions). Within the symmetric teleparallel and related theories, affine connection is metric-incompatible such that $\nabla_\mu g_{\alpha\beta}\neq0$. In order to present the formalism of symmetric teleparallel theory, one must firstly define the generalized metric affine connection \citep{PhysRevD.103.124001}:
\begin{equation}
    \Gamma^{\alpha}_{\,\,\,\mu\nu} = \widetilde{\Gamma}^{\alpha}_{\,\,\,\mu\nu}+K^{\alpha}_{\,\,\,\mu\nu}+L^{\alpha}_{\,\,\,\mu\nu}.
    \label{eq:1}
\end{equation}
In the equation above $\widetilde{\Gamma}^{\alpha}_{\,\,\,\mu\nu}$ is the usual Levi-Cevita metric-affine connection, that is widely used within the General Theory of Relativity:
\begin{equation}
    \widetilde{\Gamma}^{\alpha}_{\,\,\,\mu\nu}=\frac{1}{2}g^{\alpha\beta}(\partial_ \mu g_{\beta\nu}+\partial_\nu g_{\beta\mu}-\partial_\beta g_{\mu\nu}).
\end{equation}
While other two terms in (\ref{eq:1}) are namely contortion and deformation tensors and they could be written below as follows:
\begin{eqnarray}
    K^{\alpha}_{\,\,\,\mu\nu} &=& \frac{1}{2}g^{\alpha\beta}\left(T_{\mu\beta\nu}+T_{\nu\beta\mu}+T_{\beta\mu\nu}\right),\\
    L^{\alpha}_{\,\,\,\mu\nu} &=& -\frac{1}{2}g^{\alpha\beta}\left(Q_{\mu\beta\nu}+Q_{\nu\beta\mu}-Q_{\beta\mu\nu}\right).
\end{eqnarray}
Here $T^{\alpha}_{\,\,\,\mu\nu}=\Gamma^{\alpha}_{\,\,\,\mu\nu}-\Gamma^{\alpha}_{\,\,\,\nu\mu}$ is the torsion tensor and \citep{Capozziello:2022wgl}
\begin{equation}
    Q_{\alpha\mu\nu}=\nabla_\alpha g_{\mu\nu}=\partial_\alpha g_{\mu\nu} -\Gamma^\beta_{\,\,\,\alpha\mu}g_{\beta\nu}-\Gamma^\beta_{\,\,\,\alpha\nu}g_{\mu\beta}.
\end{equation}
Is obviously the non-metricity tensor. As we already mentioned, here and further we assume that both Ricci scalar curvature and torsion terms vanish and therefore we are left with only non-metricity. Therefore, to proceed with STEGR case one could derive the non-metricity scalar (fundamental quantity) from non-metricity tensor and its independent traces $Q_{\alpha}= {Q_{\alpha}^{\,\,\,\mu}}_{\mu}$ and $\tilde{Q}^{\alpha}=Q_{\mu}^{\,\,\,\alpha \mu}$\citep{PhysRevD.103.124001}:
\begin{equation}
    Q=-g^{\mu\nu}(L^\alpha_{\,\,\,\beta\nu}L^\beta_{\,\,\,\mu\alpha}-L^\beta_{\,\,\,\alpha\beta}L^\alpha_{\,\,\,\mu\nu})=-P^{\alpha \beta \gamma}Q_{\alpha \beta \gamma},
\end{equation}
Where deformation tensor $\mathbf{L}$ was already defined previously and superpotential could be expressed in the following way:
\begin{equation}
    P^{\alpha}_{\,\,\,\mu\nu}=\frac{1}{4}\bigg[2Q^{\alpha}_{\,\,(\mu\nu)}-Q^{\alpha}_{\,\,\,\mu\nu}+Q^\alpha g_{\mu\nu}-\delta^{\alpha}_{(\mu}Q_{\nu)}-\tilde{Q}^\alpha g_{\mu\nu}\bigg].
\end{equation}
Here symmetric and antisymmetric parts of the tensor are:
\begin{eqnarray}
    F_{(\mu\nu)}=\frac{1}{2}\bigg(F_{\mu\nu}+F_{\nu\mu}\bigg),\\
    F_{[\mu\nu]}=\frac{1}{2}\bigg(F_{\mu\nu}-F_{\nu\mu}\bigg).
\end{eqnarray}
The condition of symmetric teleparallelism makes the generic affine connection to be inertial. The most general connection is 
\begin{equation}
  \Gamma^{\alpha}_{\,\,\, \mu\nu}= \frac{\partial x^{\alpha} }{\partial \xi^{\sigma}} \frac{\partial^2 \xi^{\sigma} }{\partial x^{\mu} \partial x^{\nu}}, 
\end{equation}
where $\xi^{\sigma}$ is an arbitrary function of spacetime position. We can always  choose a coordinate $x^{\alpha}=\xi^{\sigma}$ bu utilizing a general coordinate transformation, where  the general affine  connection $\Gamma^{\alpha}_{\,\,\, \mu\nu}=0$. We call this coordinate the coincident gauge \citep{jimenez2018coincident}. Thus, in the coincident gauge, we will have $Q_{\alpha \mu \nu}= \partial_{\alpha}g_{\mu \nu}$, i.e. all the covariant derivatives are identical to ordinary derivatives.

That was the fundamentals of symmetric teleparallel analogue of the General Theory of Relativity (GR). Now we are going to present the formalism of modified symmetric teleparallel cosmology. Einstein-Hilbert action integral of the aforementioned theory of gravity is therefore could be written down as follows \citep{jimenez2018coincident}:
\begin{equation}
    \mathcal{S}[g,\Gamma,\Psi_i]=\frac{1}{16\pi G}\int_{\mathcal{M}} -d^4x e f(Q) + \mathcal{S}_{\mathrm{M}}[g,\Gamma,\Psi_i].
    \label{eq:8}         
\end{equation}
In the equation above, $\mathcal{M}$ is the four dimensional Lorentzian manifold that we work on, $g=\det g_{\mu\nu}=\prod_{\mu,\mu} g_{\mu\nu}$ is the metric tensor determinant, $e=\sqrt{-g}$ and $f(Q)$ is the arbitrary function of non-metricity scalar, that defines the modified theory of gravitation. Moreover, $\Gamma$ is the curvature free affine connection and $\mathcal{S}_{\mathrm{M}}[g,\Gamma,\Psi_i]$ defines the contribution of additional matter fields $\Psi_i$ to the total Einstein-Hilbert action integral. The reason for the above action and specific selection of the non-metricity scalar is that GR is recreated, up to a Boundary term for the choice $f=Q$, i.e., for this choice, we recover the allegedly ``symmetric teleparallel equivalent of GR". By varying the action (\ref{eq:8}) with respect to the metric tensor inverse $g^{\mu\nu}$ (using least action principle $\delta \mathcal{S}=0$) we could obtain the corresponding field equations
\begin{equation}
\frac{2}{\sqrt{-g}} \nabla_{\alpha}\left(\sqrt{-g} f_{Q} P^{\alpha \mu}_{\,\,\,\,\,\,\,\nu} \right)+ \frac{1}{2} \delta^{\mu}_{\,\,\,\nu} f + f_{Q} P^{\mu \nu \beta} Q_{\nu \alpha \beta} = T^{\mu}_{\,\,\,\,\nu},
\end{equation}
Where $f_{Q}= \frac{\partial{f}}{\partial{Q}}$ and the energy-momentum tensor could be easily derived from the variation of matter fields Lagrangian density:
\begin{equation}
    T_{\mu\nu}=-\frac{2}{\sqrt{-g}}\frac{\delta(\sqrt{-g}\mathcal{L}_\mathrm{M})}{\delta g^{\mu\nu}}.
\end{equation}
The connection of equation of motion can be computed by noticing that the variation of the connection with respect to $\xi^{\alpha}$ is equivalent to performing a diffeomorphism so that $\partial_{\xi} \Gamma^{\alpha}_{\,\, \mu\nu}=-\mathcal{L}_{\xi}  \Gamma^{\alpha}_{\,\, \mu\nu} = -\nabla_{\mu} \nabla_{\nu} \, \xi^{\alpha}$ \citep{jimenez2020cosmology}.  Besides this, in the absence of hypermomentum, one can take the variation of equation \eqref{eq:8} with respect to connection 
\begin{equation}
\label{con}
    \nabla_{\mu}\nabla_{\nu}\left( \sqrt{-g} f_{Q} P^{\mu \nu}_{\,\,\, \,\,\alpha} \right)=0.
\end{equation} 
For the metric and connection equations, one can notice that $\mathcal{D}_{\mu}T^{\mu}_{\,\,\,\,\nu}=0$, where $\mathcal{D}_{\mu}$ is the metric-covariant derivative.

Therefore, since we already defined all of the necessary quantities, we could proceed further and set up the background spacetime.

\section{FLRW cosmology}\label{sec:3}
In order to study the evolution of our universe, it will be handful to assume that background spacetime is isotropic and homogeneous, namely is Friedmann-Lemaitre-Robertson-Walker (FLRW) spacetime (we assume that lapse function is unitary):
\begin{equation}
    ds^2 = -dt^2 + \sum_{i,j}a^2(t)dx^i dx^j.
    \label{eq:11}
\end{equation}
Here $a(t)$ is the scale factor of the universe, it is a fundamental quantity that defines the evolution of the universe from its beginning. Consequently, with assumption (\ref{eq:11}), the non-metricity scalar is written as follows \citep{Caruana:2020szx}
\begin{equation}
    Q=6H^2,
\end{equation}
Where $H=\dot{a}/a$ is the well-known Hubble parameter and dot over some quantity signifies the first order temporal derivative. Finally, one could evaluate the FLRW field equations of the $f(Q)$ theory:
\begin{align}
&3H^2 = \kappa^2 \left(\rho_\text{m}+\rho_{\text{eff}}\right)\, , \\
&3H^2 + 2\dot{H}= -\kappa^2\left(p_\text{m}+p_{\text{eff}}\right)\, . 
\end{align}
Here $\kappa^2=8\pi G$ is the Einstein gravitational constant squared, $\rho_\text{m}$ and $p_\text{m}$ are, respectively, matter energy density and isotropic pressure. Moreover, in the equation above $\rho_{\text{eff}}$ and $p_{\text{eff}}$ are effective energy density and pressure that define the contribution of $f(Q)$ gravity to the field equations. For modified STEGR, fields equations with plugged exact forms of effective quantities read:
\begin{equation}
    3H^2 = \frac{\kappa^2}{2f_Q}\bigg(\rho_m+\frac{f}{2}\bigg),
    \label{eq:17}
\end{equation}
\begin{equation}
    \left(12 H^{2} f_{QQ} + f_{Q}\right) \dot{H} = -\frac{k^{2}}{2} \left( \rho_{m} + p_{m} \right).
    \label{eq:18}
\end{equation}
are effective DE energy density and pressure that define the contribution of the $f(Q)$ cosmology with
\begin{equation}
    f_Q=\frac{\partial f(Q)}{\partial Q},\quad f_{QQ}=\frac{\partial^2 f(Q)}{\partial Q^2}
\end{equation}

The energy-momentum tensor of the cosmological fluid which is given by
\begin{equation}
T_{\mu \nu}= \left( \rho + p  \right) u_{\mu}u_{\nu} + pg_{\mu \nu},
\end{equation}
which leads to conservation equation as $\dot{\rho}+ 3 H \left(\rho+p\right)=0$.
In symmetric teleparallel gravity and its extensions, the conservation law $T^{\mu}_{\,\,\,\nu;\mu}=0$ holds for the matter energy-momentum tensor. The $T^{\mu}_{\,\,\,\nu;\mu}=0$ holds through \eqref{con} for the connection \citep{jimenez2018teleparallel,dimakis2022flrw,harko2018coupling}. 

\subsection{Exponential $f(Q)$ gravity}
This paper is particularly aimed at the investigation of one $f(Q)$ gravity model - namely modified exponential $f(Q)$ gravity (which is built from the linear and exponential terms respectively). In $f(Q)$ theory, numerous cosmic possibilities have been examined using various exponential models, notably inflationary cosmology, BBN constraints, and dynamic system analysis.\citep{harko2018coupling,anagnostopoulos2023new,khyllep2023cosmology}. For that kind of gravity, $f(Q)$ function reads (we adapt the work of \citep{linder2009exponential,PhysRevD.81.127301} for modified STEGR):
\begin{equation}
    f(Q) = Q+\alpha Q_0(1-e^{-\beta\sqrt{Q/Q_0}})
\end{equation}
Where $\alpha$, $\beta$ are free MOG parameters, namely additional degrees of freedom. We can reduce the number of d.o.f by matching first Friedmann equation (\ref{eq:17}) at the present time (i.e. assuming that $z=0$):
\begin{equation}
\alpha = -\frac{e^\beta(-1+\Omega_{m0}+\Omega_{r0})}{-1+e^\beta-\beta}    
\end{equation}
Thus, the complexity of this form is just one step more than the standard $\Lambda$CDM. The exponential modified gravity could satisfy all stability and validity and not cross the pahntom divide line \citep{arora2022crossing}. In order to solve the field equations and obtain the numerical form of Hubble parameter, the definition of $\dot{H}$ is 
\begin{equation}
    \dot{H}=aH\frac{dH}{da}.
\end{equation}
We will solve the aforementioned  field equation numerically, as we already stated, with \texttt{Mathematica} numerical ODE solver \texttt{NDSolve}. Initial conditions at the vanishing redshift for $\dot{H}$ could be therefore set up (as a cosmographical quantity)\citep{Mandal2020}:
\begin{equation}
\dot{H}_0=-H_0^2(1+q_0).
\end{equation}
Where $q_0$ is the current deceleration parameter, we fix it to $q_0=-0.55$ \citep{Reid:2019tiq}. Additionally, for MCMC training, as a truths we assume that the present value of the Hubble parameter is $H_0=69 \, \mathrm{km/s/Mpc}$ and that matter mass fraction at the present time is $\Omega_{m0}=0.315\pm 0.007$, following the observational constraints of \textit{Planck2018} \citep{2020A&A...641A...6P}.

\section{MCMC constraints}\label{sec:4}
In this section, we want to constrain our $f(Q)$ gravity model via observational datasets. To explore the parameter space, we will be using the Markov Chain Monte Carlo (MCMC) methodology and Python package \texttt{emcee} \citep{2013PASP..125..306F}.

\subsection{Observational Hubble Data}

Observational Hubble Data is one of the most popular and plausible tests of the universe expansion history beyond GR and $\Lambda$CDM. The OHD sample is mainly obtained from the differential age of galaxies method (or just DAG) \citep{Yu_2018,moresco2015raising}. In this method, Hubble rate is usually obtained from the formula below
\begin{equation}
  H(z)=\frac{-1}{1+z}\frac{dz}{dt}
\end{equation}
In the current article, we will primarily use the OHD points derived from the so-called Cosmic Chronometers (CC)  i.e., the massive and passively evolving galaxies. Using Cosmic Chronometers, ratio $dz/dt$ could be derived from the $\Delta z/\Delta t$, where $\Delta z$ is the redshift separation in the galaxies sample and could be easily determined through precise and accurate spectroscopy. On the other hand, derivation of the $\Delta t$ is much more challenging and requires some standard clocks. For that purpose, we could use massive, passively evolving, and old stellar populations that are present across a wide range of redshifts and therefore could be considered as cosmic chronometers.
 To determine the priors and likelihood functions (which are necessary), we used the $H(z)$ dataset. To constrain our modified gravity model, we introduce the chi-squared function below
\begin{equation}
    \chi^2_{CC}=\sum_{i=1}^{N_H}\bigg[\frac{H^\mathrm{th}_i(p_1,p_2,...,p_n,z_i)-H^\mathrm{obs}_i(z_i)}{\sigma_{H(z_i)}}\bigg]^2
\end{equation}
The likelihood function that we will be using for MCMC sampling have its usual exponential form
\begin{equation}
    \mathcal{L}=\exp(-\chi^2/2)
\end{equation}

\subsection{Pantheon SN Ia Sample}
We also used the Pantheon dataset to constrain our modified gravity with dark energy, which consists of data obtained from the 1048 Ia supernovae (discovered by the PANSTARRS DR1 (PS1) Medium Deep
Survey, Low $z$, SNLS, SDSS and HST \citep{Scolnic_2018,Chang_2019}). For this dataset, redshift varies from $z=0.01$ to $z=2.26$.
The corresponding chi-squared function reads:
\begin{equation}
    \chi^2_{SN}(p_1,p_2,...,p_n)= \sum^{N_{SN}}_{i,j=1}\frac{\Delta \mu_i}{\sigma_{\mu(z_i)}}
\end{equation}
Where
\begin{equation}
    \Delta\mu_i=\mu^{th}(p_1,p_2,...,p_n)-\mu_i^{obs}
\end{equation}
And distance moduli is \citep{Arora_2020b}
\begin{equation}
    \mu^{th}=5\log_{10}D_L(z)+\mu_0,\quad \mu_0 = 5 \log_{10} \frac{H_0^{-1}}{\mathrm{Mpc}}+25
\end{equation}
\begin{equation}
    D_L(z)=\frac{c(1+z)}{H_0}S_K\bigg(H_0\int^z_0\frac{d\overline{z}}{H(\overline{z})}\bigg)
\end{equation}
Here function $S_K(x)$ is just
\begin{equation}
S_K(x)=    \begin{cases}
      \sinh(x\sqrt{\Omega_K})/\Omega_K,\quad \Omega_K >0\\
      x,\quad\quad\quad\quad\quad\quad\quad\;\;\; \Omega_K=0\\
      \sin (x \sqrt{|\Omega_K|})/|\Omega_K|,\quad \Omega_K<0
    \end{cases}\,.
\end{equation}
It is known that out universe is spatially flat, and therefore $\Omega_K=0$. \\
The nuisance parameters in the Tripp formula \citep{tripp1998two} $\mu= m_{B}-M_{B}+\alpha x_{1}-\beta c+ \Delta_{M}+\Delta_{B}$ were retrieved using the novel method known as BEAMS with Bias Correction (BBC) \citep{kessler2017correcting}, and the observed distance modulus is now equal to the difference between the corrected apparent magnitude $M_{B}$ and the absolute magnitude $m_{B}$ $\left(\mu= m_{B}-M_{B}\right)$. Additionally, one can define the chi-squared function in terms of covariance matrix as follows \citep{Deng:2018jrp}:
\begin{equation}
\chi^2_{\mathrm{SN}}=\Delta \boldsymbol{\mu}^T \mathbf{C}^{-1} \Delta \boldsymbol{\mu}
\end{equation}
Where covariance matrix consists of systematic and statistical uncertainties respectively \citep{2011ApJS..192....1C}:
\begin{equation}
\mathbf{C} = \mathbf{D}_{\mathrm{stat}}+\mathbf{C}_{\mathrm{sys}}
\end{equation}
In the current work we assume that diagonal matrix of statistical uncertainties looks like $\mathbf{D}_{\mathrm{stat},ii}=\sigma^2_{\mu(z_i)}$. Besides, systematic uncertainties are derived using the Bias Corrections (BBC) method, introduced and developed in \citep{Pan-STARRS1:2017jku}:
\begin{equation}
\mathbf{C}_{ij,\mathrm{sys}} = \sum^K_{k=1}\bigg(\frac{\partial \mu^{obs}_i}{\partial S_k}\bigg)
\bigg(\frac{\partial \mu^{obs}_j}{\partial S_k}\bigg)\sigma^2_{S_k}
\end{equation}
Indexes $\{i,j\}$ denote the redshift bins for distance modulus, $S_k$ here denotes the magnitude of systematic error, $\sigma_{S_k}$ is respectively it's standard deviation uncertainty.
\begin{figure*}
    \centering
    \includegraphics[width=\textwidth]{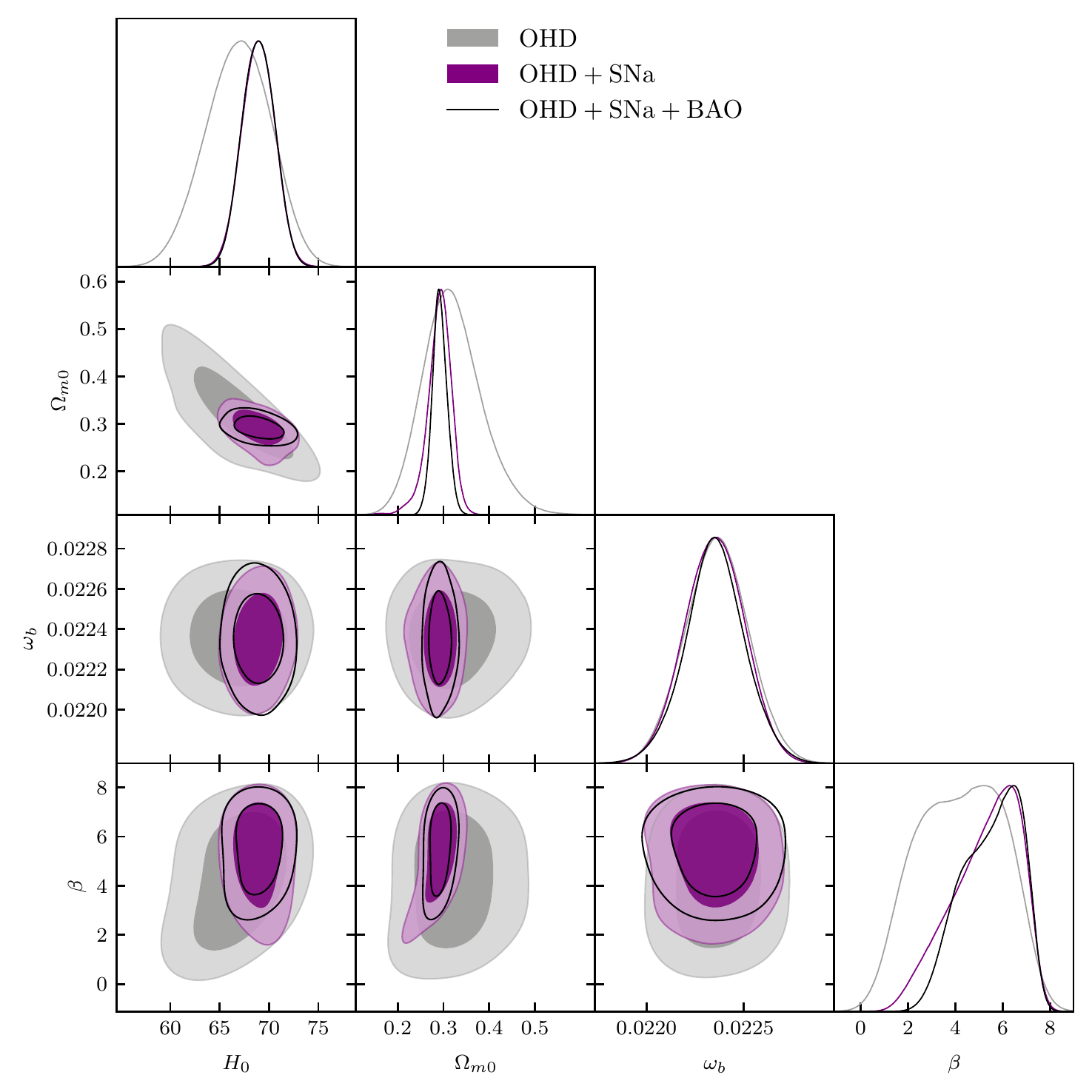}
    \caption{MCMC best fits from OHD, Pantheon and BAO datasets and joint distribution for exponential $f(Q)$ model}
    \label{fig:1}
\end{figure*}
\subsection{Baryon Acoustic Oscillations}
Finally, we use Baryon Acoustic Oscillations (BAOs) to constrain our modified gravity model. BAOs arise in the early times of universe evolution. In earlier times, fermions and photons are strongly connected to each other due to the Thompson scattering. This mixture of baryons and photons behaves like a single fluid and can not gravitationally collapse. Moreover, this fluid oscillates because of the huge photonic pressure. These oscillations are called BAOs. The Characteristic scale of the BAO is defined by the so-called sound horizon $r_s$, which is seen at the photon decoupling epoch with redshift $z_*$:
\begin{equation}
    r_s=\frac{c}{\sqrt{3}}\int^{\frac{1}{1+z_*}}_0\frac{da}{a^2 H\sqrt{1+(3\Omega_{b0}/4\Omega_{\gamma0})a}}
\end{equation}
Here $\Omega_{b0}$ is known as the baryon mass density at present ($z=0$) and $\Omega_{\gamma0}$ is, respectively, photon mass density at present. Also, as it was noticed, angular diameter distance is derived directly from the BAO sound horizon. In this work, to constrain our MOG model with BAO, we will use the observational datasets with $d_A(z_*)/D_V(z_{BAO})$ data. Here we consider that $d_A(z_*)$ is the angular diameter distance in the comoving coordinates:
\begin{equation}
    d_A(z)=\int ^z_0\frac{dz'}{H(z')}
\end{equation}
And $D_V(z_{BAO})$ is the dilation scale:
\begin{equation}
    D_V(z)=(d_A(z)^2z/H(z))^{1/3}
\end{equation}
Finally, we also consider that photon decoupling epoch arise at the redshift \citep{2016A&A...594A..13P}:
\begin{equation}
    z_* = 1048[1+0.00124(\Omega_bh^2)^{-0.738}][1+g_1(\Omega_mh^2)^{g_2}]
\end{equation}
Where,
\begin{equation}
    g_1=\frac{0.0783(\Omega_bh^2)^{-0.238}}{1+39.5(\Omega_bh^2)^{-0.763}}
\end{equation}
\begin{equation}
    g_2=\frac{0.560}{1+21.1(\Omega_bh^2)^{-1.81}}
\end{equation}
This dataset was gathered from the works of \citep{10.1111/j.1365-2966.2011.19592.x,10.1111/j.1365-2966.2009.15812.x,Jarosik_2011,Eisenstein_2005,Giostri_2012}.

Consequently, to perform the MCMC sampling, we need to define the chi squared function for our BAO dataset:
\begin{equation}
    \chi^2_{BAO}=X^T C^{-1}X
\end{equation}
Where $X$ is the matrix of form \citep{Giostri_2012}:
\begin{equation}
X=\left( 
\begin{array}{c}
\frac{d_{A}(z_{* })}{D_{V}(0.106)}-30.95 \\ 
\frac{d_{A}(z_{* })}{D_{V}(0.2)}-17.55 \\ 
\frac{d_{A}(z_{* })}{D_{V}(0.35)}-10.11 \\ 
\frac{d_{A}(z_{* })}{D_{V}(0.44)}-8.44 \\ 
\frac{d_{A}(z_{* })}{D_{V}(0.6)}-6.69 \\ 
\frac{d_{A}(z_{* })}{D_{V}(0.73)}-5.45%
\end{array}%
\right) \,
\end{equation}
 
We also performed the joint analysis form the combined $OHD+SN+BAO$ by minimizing $\chi^{2}_{OHD} +\chi^{2}_{SN} +\chi^{2}_{BAO}$.

\begin{table}
\begin{center}
\label{table1}
\adjustbox{width=0.5\textwidth}{
\begin{tabular}{l c c c c}
\noalign{\doubleline}
Datasets& $H_0$ & $\Omega_{m0}$ & $\beta$ \\
\hline
Hubble (OHD)& $66.9\pm3.3$ & $0.320^{+0.055}_{-0.070}$   & $4.3\pm1.9$   \\ 
OHD+SNa   & $68.9\pm1.7$ & $0.290^{+0.028}_{-0.020}$   & $5.3^{+1.8}_{-1.0}$   \\ 
OHD+SNa+BAO   & $68.9\pm1.6$ & $0.292\pm0.016$   & $5.6\pm 1.25$   \\ 
\hline 
$\rm Models$ & $\chi_{\rm min}^2$ & $\rm AIC$ & $\rm BIC$ \\
\hline 
$\Lambda$CDM & 58.700 & 67.248 & 76.127     \\ 
$f(Q)$   & 57.616 &  68.499 &  79.137  \\
\hline
\end{tabular}
}
\caption{Best-fit values of model parameters and statistical analysis}
\end{center}
\end{table}

The results are, therefore, numerically derived from MCMC trained on OHD, Pantheon, BAO and joint datasets. Besides, results are placed on the Table (\ref{table1}) above for model free parameters $H_0$, $\beta$ and $\Omega_{m0}$. Furthermore, the $1-\sigma$ and $2-\sigma$ likelihood contours for the possible subsets of parameter space  are presented in Fig. \ref{fig:1}.

\subsection{Statistical evaluation}
To evaluate the success of our MCMC analysis, one should perform the statistical evaluation using the so-called Akaike Information Criterion (AIC) and Bayesian Information Criterion (BIC). The first quantity, namely AIC can be expressed as follows \citep{1100705}:
\begin{equation}
    \mathrm{AIC} = \chi^2_{\mathrm{min}}+2d
\end{equation}
With $d$ being the number of free parameters in a chosen model. To compare our results with the well-known fiducial $\Lambda$CDM model, we are going to use the AIC difference between our modified gravity model and fiducial cosmology $\Delta\mathrm{AIC}=|\mathrm{AIC}_{\Lambda\mathrm{CDM}}-\mathrm{AIC}_{\mathrm{MOG}}|
$. In that case, if $\Delta\mathrm{AIC}<2$, there is a strong evidence in favor of MOG model, while for $4<\Delta\mathrm{AIC}\leq 7$ there is a little evidence if favor of MOG model of our consideration. Finally, for the case with $\Delta \mathrm{AIC}>10$ there is practically no evidence in favor of MOG \citep{10.1111/j.1745-3933.2007.00306.x}. In addition, BIC is defined through the relation, written down below:
\begin{equation}
    \mathrm{BIC} =\chi^2_{\mathrm{min}}+d\ln N
\end{equation}
For that case, $N$ is the number of data points being used for MCMC. For BIC, if $\Delta \mathrm{BIC}<2$, there is no strong evidence against chosen model that deviate from $\Lambda$CDM, if $2\leq \Delta \mathrm{BIC}<6$ there is evidence against the MOG model and finally for $\Delta\mathrm{BIC}>6$ there is strong evidence against MOG model. We therefore store the $\chi^2_{\mathrm{min}}$/AIC/BIC data for modified gravity model of our consideration in the Table (\ref{table1}). As we see, $\Delta\rm AIC=1.25$ and $\Delta\rm BIC=3.01$ so that our model can very precisely mimic $\Lambda$CDM one.
\section{Validity of cosmological constraints}\label{sec:5}
\begin{figure}
    \centering
    \includegraphics[width=\columnwidth]{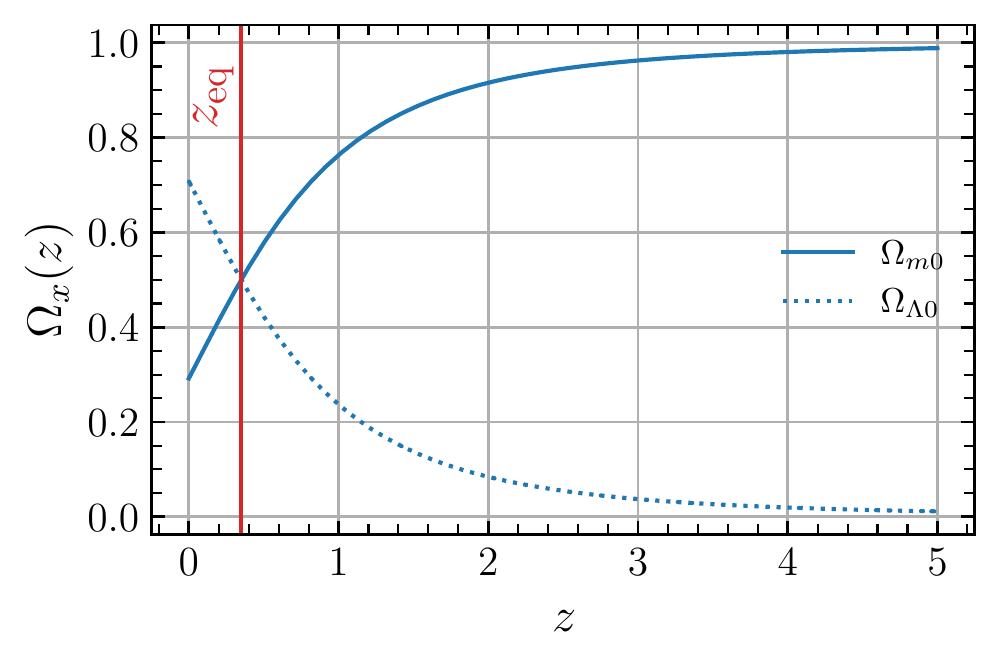}
    \caption{Dimensionless mass density for matter and effective dark energy within exponential $f(Q)$ gravitation}
    \label{fig:omega}
\end{figure}
\begin{figure}
    \centering
    \includegraphics[width=\columnwidth]{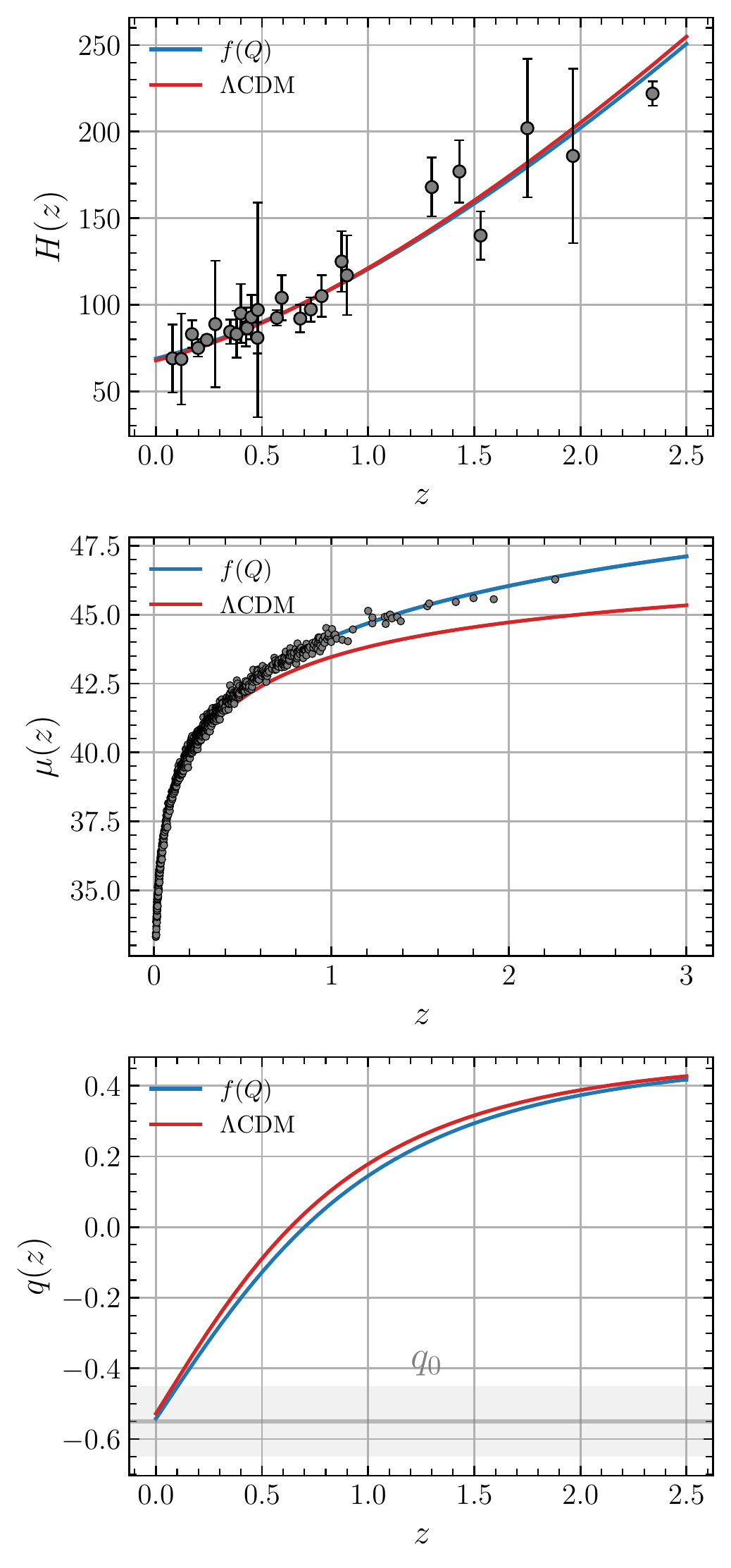}
    \caption{Hubble parameter $H(z)$, deceleration parameter $q(z)$ and distance modulus for exponential $f(Q)$ gravity with best fit values from MCMC used. For comparison, we as well show the fiducial $\Lambda$CDM results}
    \label{fig:2}
\end{figure}
In order to check the validity of the aforementioned applied cosmological constraints (such as Pantheon, BAO, or OHD), we will probe the behavior of some quantities, such as deceleration parameter $q(z)$ or statefinder pair. Furthermore, one could define the so-called deceleration parameter:
\begin{equation}
    q=-\frac{\dot{H}}{H^2}-1.
\end{equation}
From the above equation, one could easily notice that both deceleration and Hubble parameters are related to each other by higher order temporal derivatives of scale factor $a$. Consequently, to differentiate our model from other numerous MOG, DE/DM models, one could present the pair of parameters, the so-called statefinder pair \citep{sahni2003statefinder,alam2003exploring,Pasqua2015,Xu:2017qxf}:
\begin{equation}
    r = \frac{\dddot{a}}{aH^3},
\end{equation}
\begin{equation}
    s=\frac{r-1}{3(q-1/2)}.
\end{equation}
For the sake of simplicity, we could redefine the statefinder pair $\{r,s\}$ fully in terms of deceleration parameter:
\begin{equation}
    r(z)=q(z)(1+2q(z))+q'(z)(1+z),
\end{equation}
\begin{equation}
    s(z)=\frac{r(z)-1}{3(q(z)-1/2)}.
\end{equation}
We are going to construct the phase plane $r(z)-s(z)$, in which different points correspond to the various universe states, such that:
\begin{itemize}
    \item $\Lambda$CDM corresponds to $(s=0,r=1)$,
    \item Chaplygin Gas (CG) corresponds to $(s<0,r>1)$,
    \item SCDM corresponds to $(r=1,q=0.5)$,
    \item Quintessence corresponds to $(s>0,r<1)$.
\end{itemize}

\begin{figure}
    \centering
    \includegraphics[width=\columnwidth]{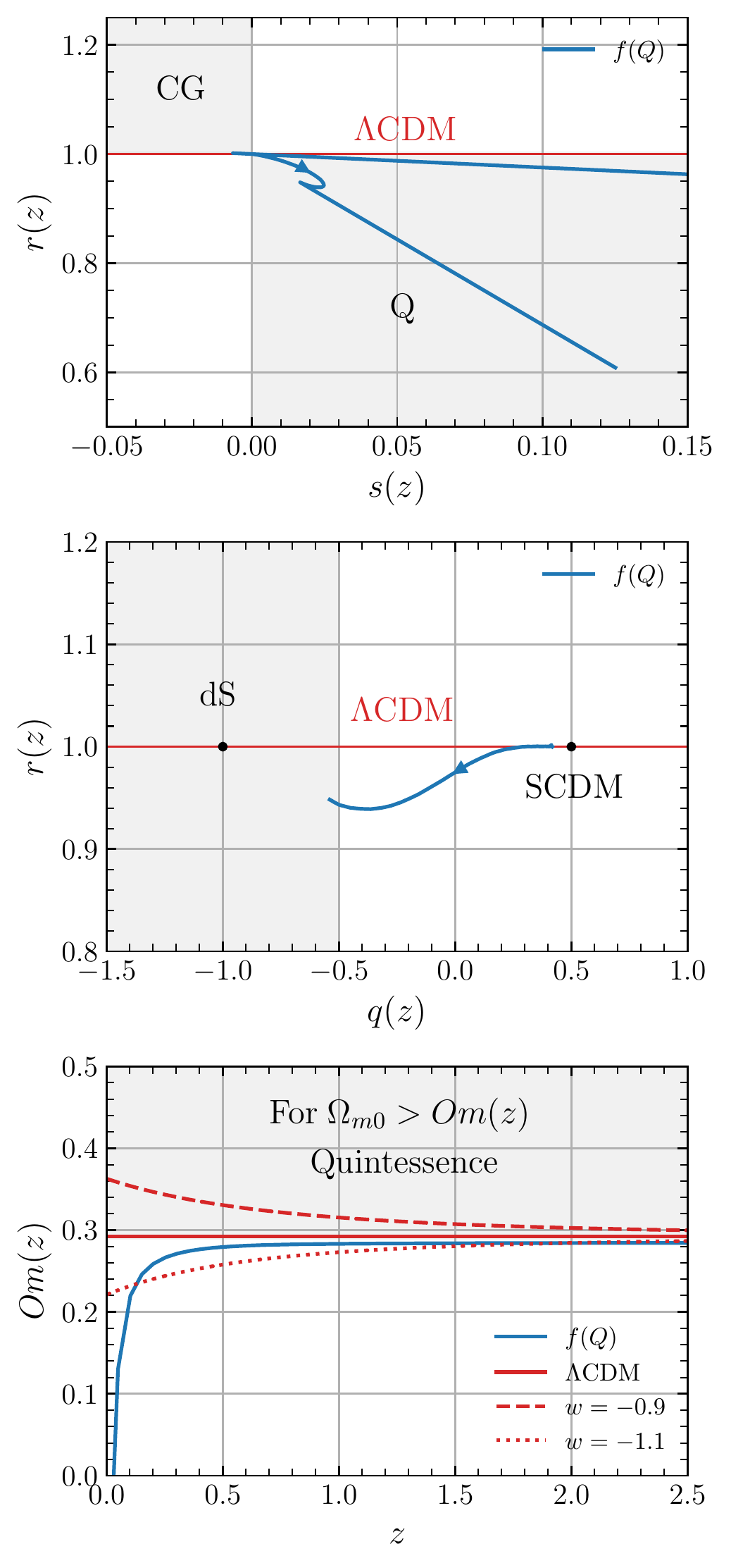}
    \caption{Statefinder pairs and $Om(z)$ function for exponential $f(Q)$ gravity, $\Lambda$CDM and $\omega$ varying $\Lambda$CDM cosmologies}
    \label{fig:3}
\end{figure}

Consequently, we plot both statefinder parameter phase portraits, deceleration parameter, and additionally $H(z)$ as probes of model validity in the cosmological sense in Figs. (\ref{fig:2}) and (\ref{fig:3}). Statefinder diagnostics and $q(z)$ were performed only for the joint dataset, since other datasets shows the similar behavior as the joint solution. Remarkably, a transition from deceleration to acceleration phase on the third plot of the aforementioned figure is seen. A valid interval for $q_{0}$ is marked as a gray area. As stated already, one may check the universe evolutionary scenario using statefinder pairs $\{r,s\}$ and $\{r,q\}$. From the $r-s$ plane of our model, one could observe that the initial universe was filled with quintessence, then passed the $\Lambda$CDM phase and is currently reverting towards the quintessence scenario. On the other hand, in the $r-q$ plane, it is evident that the universe once also passed through the $\Lambda$CDM phase. However, now our space-time is generally filled with quintessential fluid, it is expected that the future universe will eventually turn to the de-Sitter state (when $\Lambda$ term will fully dominate). The point on quintessential fluid also coincides with MCMC observational constraints.

The very last probe of cosmological validity is the well known $Om(z)$ diagnostics, firstly presented in the paper \citep{PhysRevD.78.103502}, where $Om(z)$ is defined through equation \citep{10.1093/mnras/sty755,https://doi.org/10.48550/arxiv.2203.08907}:
\begin{equation}
    Om(z)=\frac{E^2(z)-1}{(1+z)^3-1}
    \label{eq:46}
\end{equation}
This parameter was derived to distinguish $\Lambda$CDM from other, more complicated cosmological models. In the equation (\ref{eq:46}), it is a handful to define $E^2(z)=H^2(z)/H_0^2$, which is exactly the Hubble flow, dimensionless quantity normalized by the current Hubble parameter value.  $Om(z)$ parameter have a constant value for the $\Lambda$CDM model, which is same as the current matter mass density $\Omega_{m0}$.

Consequently, we place numerical solution of $Om(z)$ function for $f(Q)$ model in the Fig.(\ref{fig:3}). For the sake of comparison, we as well plot $Om(z)$ solutions within classical $\Lambda$CDM model and within $\omega$ varying $\Lambda$CDM cosmologies. As one could easily notice, for our $f(Q)$ model, $Om(z)$ shows only $Om(z)<\Omega_{m0}$ behavior in the distant past, which could lead to the presence of phantom fluid (for more information on the subject, see \citep{Mostaghel:2016lcd}). 
\begin{figure*}
    \centering
    \includegraphics[width=\textwidth]{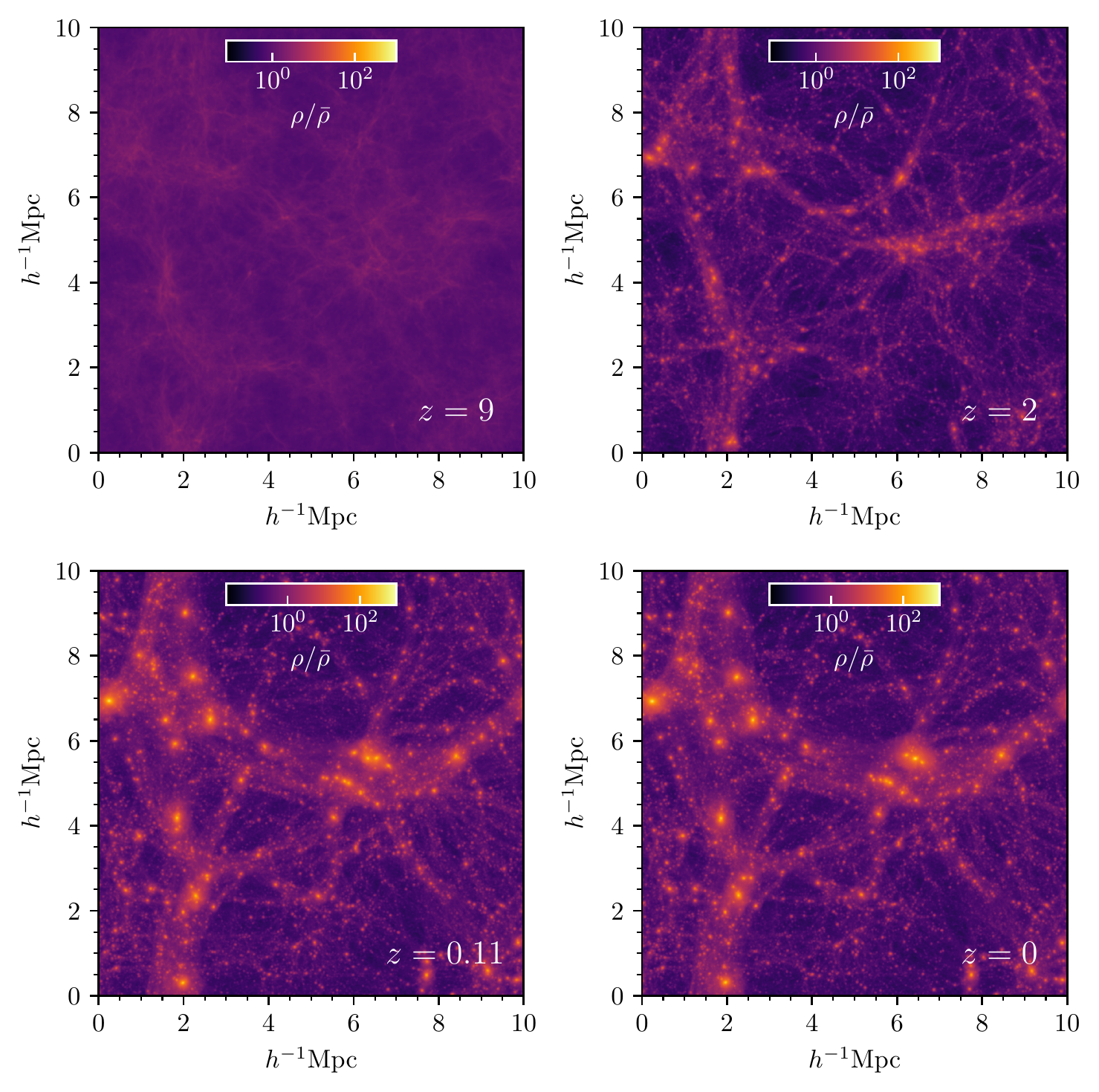}
    \caption{N-body simulations snapshot (CDM over density) for $f(Q)$ gravity with best fit MCMC values on different redshifts}
    \label{fig:66}
\end{figure*}
\begin{figure*}
    \centering
    \includegraphics[width=\textwidth]{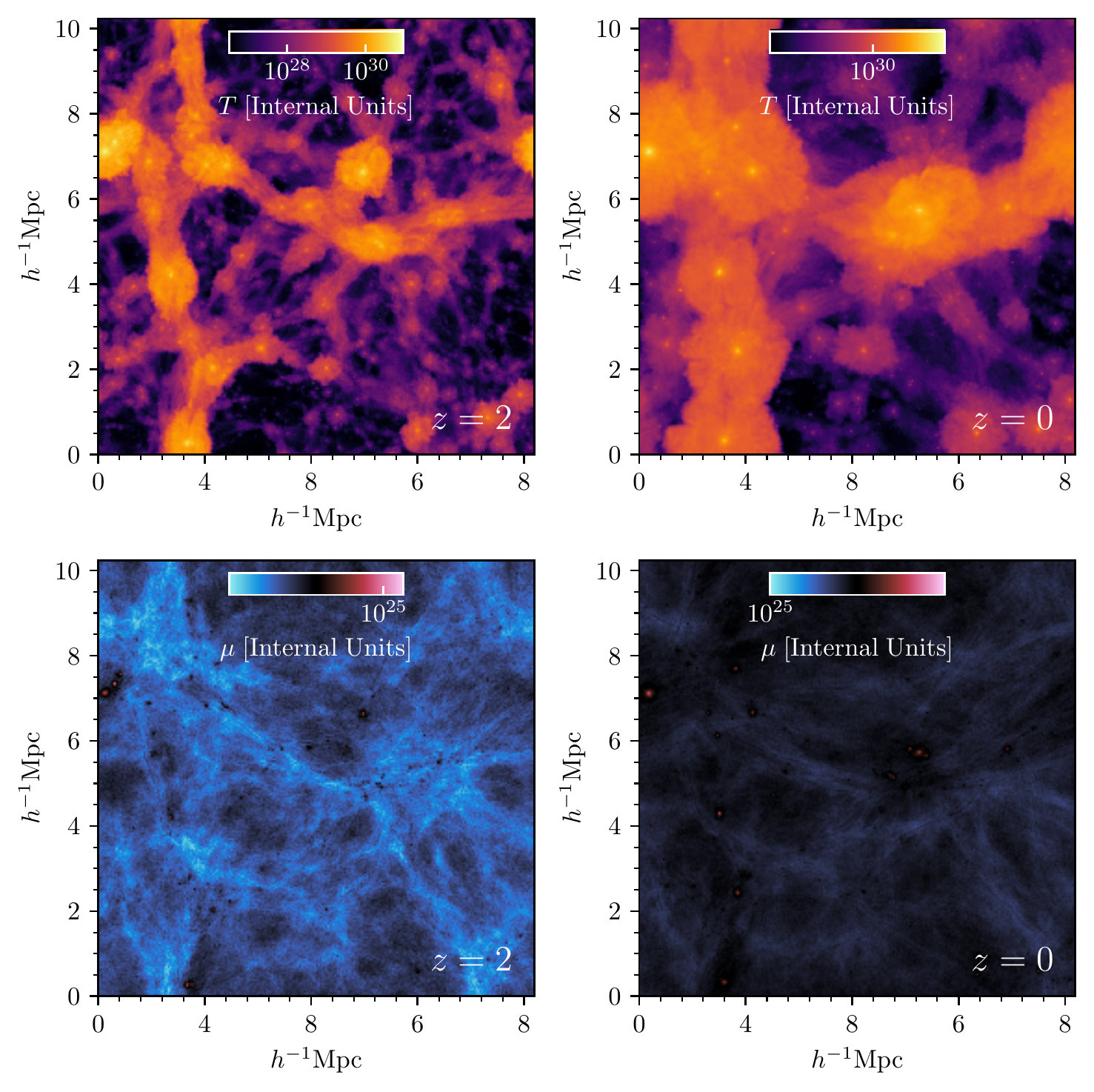}
    \caption{SPH simulation snapshots of $f(Q)$ gravity for gas temperature $T$ and mean molecular weight $\mu$}
    \label{fig:77}
\end{figure*}
However, at $z\approx2$, our model transits $\Lambda$CDM and has constantly growing trend, therefore in the near past and present times, quintessential fluid appears, which converges well with the statefinder diagnostics and MCMC. 

Finally, we also analyse both matter and effective dark energy mass densities for our model of modified gravitation to conclude on its validity. Corresponding results are plotted on the Figure (\ref{fig:omega}). One can easily notice, both $\Omega_{m0}\land\Omega_{\Lambda 0}\in[0,1]$ and their sum always converges to unity, epoch of the equality appears at redshift $z\approx 0.35$, which is very near to the $\Lambda$CDM estimate.

\section{N-body simulations of LSS with small $\mathrm{L_{\mathrm{Box}}}$}\label{sec:6}
As we already remarked previously, the main purpose of this paper is to perform N-body simulations of the comoving box that contain DM+baryonic matter and dark energy in exponential $f(Q)$ gravitation and compare our results with the Large Scale Structure of concordance $\Lambda$CDM cosmology. For that aim, we will use the publicly available code \texttt{ME-GADGET}, a modification of the well-known hydrodynamical N-body code \texttt{GADGET2}. It was modified for generality so that one can perform simulations for practically any cosmological model. The code above was described in the pioneering works of \citep{2019MNRAS.489..297A,2019ApJ...875L..11Z}, whereas the tests are provided in \citep{2018PhRvD..98j3530Z}. This code as an input needs tables with Hubble flow $H/H_0$ and the deviation of effective gravitational constant from the Newtonian one $G_{\mathrm{eff}}/G_N$ (in some models of modified gravity, namely screened ones, such deviation exists only up to some scale $k_{\mathrm{screen}}$ because of the so-called fifth force). The effective gravitational constant exact form was found in the paper \citep{PhysRevD.101.103507}:
\begin{equation}
    G_{\mathrm{eff}}=\frac{G_N}{f_Q}
\end{equation}
Equation above is being numerically solved assuming appropriate best fit values for free parameters of our model. As one could easily notice, at the very early time (high-$z$ epochs), $f(Q)$ gravity has Newtonian-like gravitational constant and then, at approximately $a\approx0.1$, $G_{\mathrm{eff}}$ is being separated from $G_{\mathrm{N}}$ for our model.

Since we already defined needed inputs for \texttt{ME-GADGET} code, we could proceed further to fine-tuning our simulation setup.

\subsection{Simulation setup}
One needs to define various parameters to produce the simulations and initial conditions, based on the second order Lagrangian Perturbation Theory (namely, 2LPT). We want to obtain the mid-resolution simulations, therefore, particle number is $N=512^3$ and mesh size is respectively $N_{\mathrm{mesh}}=2\times512^3$ as well. The simulation box has periodic vacuum boundary conditions and sides with length $10\mathrm{Mpc/h}$. Initial conditions were produced with the \texttt{Simp2LPTic} code (see GitHub repository \href{https://github.com/liambx/Simp2LPTic}{https://github.com/liambx/Simp2LPTic}), glass files (pre-initial conditions) were generated with the use of \texttt{ccvt-preic} (check \href{https://github.com/liaoshong/ccvt-preic}{https://github.com/liaoshong/ccvt-preic}). We assumed that glass tile fraction is unitary. Moreover, cosmological parameters were borrowed from our MCMC constraints, discussed earlier: $h=100H_0=0.689\pm 0.016$ (so-called "little-h"), $\Omega_{m0}=0.292 \pm 0.016$, leading to $\Omega_{\Lambda0}=0.708$, if one will not take into account radiation and massive neutrino species. On the other hand, baryon mass density equals to $\Omega_b=0.0493$ (relation between total matter density and baryon one decides how much gas particles are present in the simulation). Moreover, matter power spectrum amplitude at $k=8\mathrm{Mpc/h}$ is assumed to be $\sigma_8=0.811 \pm 0.006$ and initial power spectrum is linear, constructed from the Eisenstein \& Hu transfer function \citep{1998ApJ...496..605E} (power spectrum were constructed using code \texttt{CAMB}, see \citep{2011ascl.soft02026L}). Initial conditions are being generated at the redshift $z=10$ and spectrum index of scalar perturbations is $n_s=0.9649\pm 0.0042$ \citep{PlanckInf}.
\subsection{Results}
In the current subsection, we are going to discuss the main results, obtained from the N-body simulations of the Large Scale Structure of the Universe.

Firstly, we demonstrate the spatial slices of CDM over density $\delta_{\mathrm{CDM}}=\overline{\rho}_{\mathrm{CDM}}/\rho_{\mathrm{CDM}}$ for our $f(Q)$ gravity model with different values of redshift $z$ on the Figure (\ref{fig:66}).

In addition to the over density measurements, we also show the temperature of gas $T$ that arise from Smoothed Particle Hydrodynamics (SPH) and mean molecular weight $\mu=\overline{m}/m_{\mathrm{HI}}$, which defines the relation between mean particle mass and neutral hydrogen particle mass on the Figure (\ref{fig:77}) respectively. As one can easily notice, the DM walls are represented by smaller value of mean molecular weight. Besides, temperature maps show the well-known hot "bubbles" within the Inter-Galactic Medium (IGM) that are formed due to impinging galactic winds.

Now we are going to investigate the matter power spectrum for our model. In comparison, we are going to use fiducial $\Lambda$CDM cosmology power spectrum, generated with the use of \texttt{CAMB} code \citep{2011ascl.soft02026L,Lewis:2002ah,Lewis:1999bs,Howlett:2012mh}\footnote[1]{Documentation for this code is stored in \href{https://camb.readthedocs.io/en/latest/}{camb.readthedocs.io}}. In order to extract $P(k)$ for some value of redshift within our N-body framework, we used Python-based code \texttt{Pylians3} \citep{Pylians}\footnote[2]{For installation procedure and full documentation, refer to the \href{https://pylians3.readthedocs.io/en/master/}{pylians3.readthedocs.io}}.

We consequently compare the matter power spectrum on the Figure (\ref{fig:77}) with/without Redshift-Space Distortions (RSDs) directed along both $X$, $Y$ and $Z$ axes. As we noticed during numerical analysis, up to some $k$ near $k_{\mathrm{Box}}$ limit for our simulation, $P(k)$ spectrum in Fourier space does reconstruct non-linear matter power spectrum, given by \texttt{CAMB}, while Redshift-Space Distorted (RSD) one behave like linear matter power spectrum, as expected. Also, it is worth to notice that difference between RSD and regular matter power spectrum is bigger for CDM+Gas case. Finally, effect of RSDs in our simulations is almost isotropic, so that $\Delta(\mathrm{RSD})$ differs only by few percents with the change of RSD direction axis.
\begin{figure}
    \centering
    \includegraphics[width=\columnwidth]{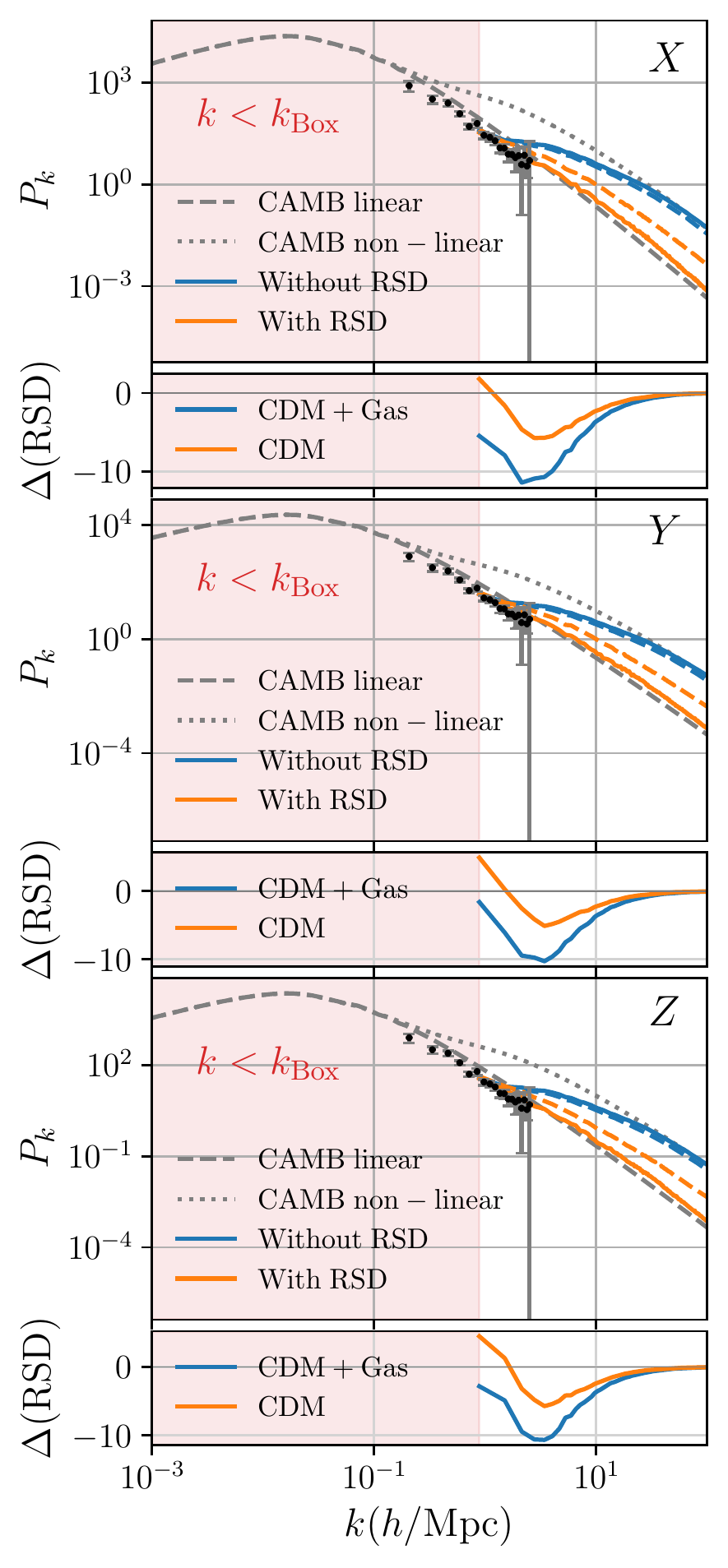}
    \caption{Matter power spectrum with/without RSDs for $f(Q)$ gravity vs. \texttt{CAMB} linear/non-linear $P(k)$ for $\Lambda$CDM. Dashed N-body $P(k)$ represent the CDM-only power spectrum, while solid line represent CDM+Gas $P(k)$. Error bars represent Ly $\alpha$ forest observations on high $z$}
    \label{fig:77}
\end{figure}
\begin{figure}
    \centering
    \includegraphics[width=\columnwidth]{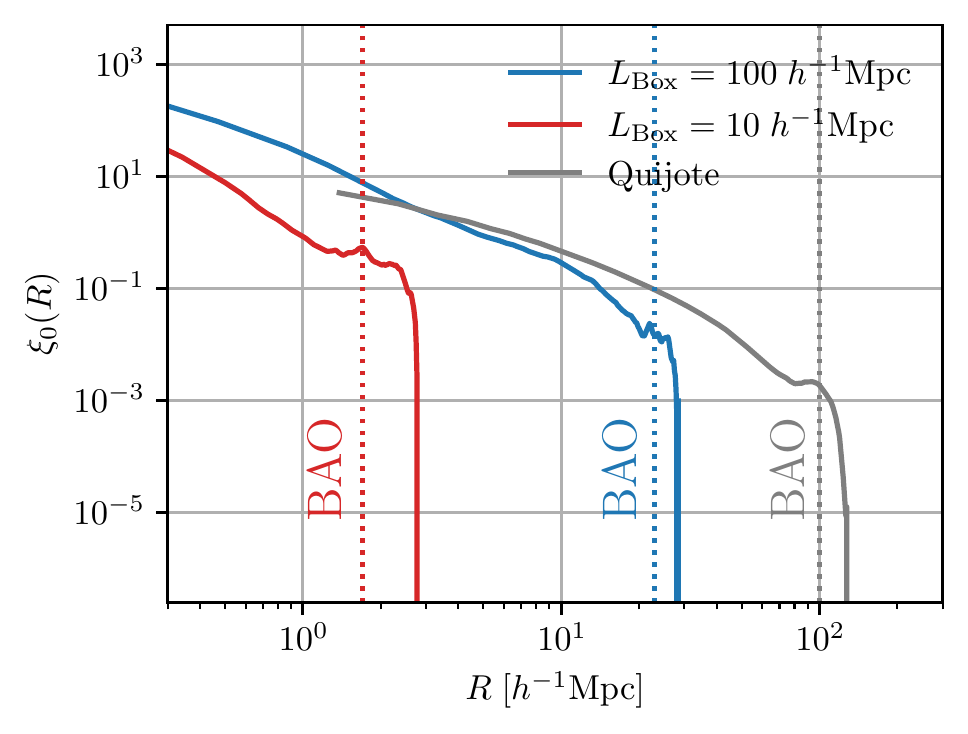}
    \caption{Monopole redshift-space distorted two point correlation function with $L_{\mathrm{Box}}=10/100h^{-1}$Mpc for $f(Q)$ log-like modified gravity. In relation we plot Quijote simulation correlation function for Planck fiducial cosmology with Gpc wide box. Also, for each case we display the scale, at which BAO bump occurs}
    \label{fig:corrfunc}
\end{figure}
\subsection{Halo mass function}
\begin{figure}
    \centering
    \includegraphics[width=\columnwidth]{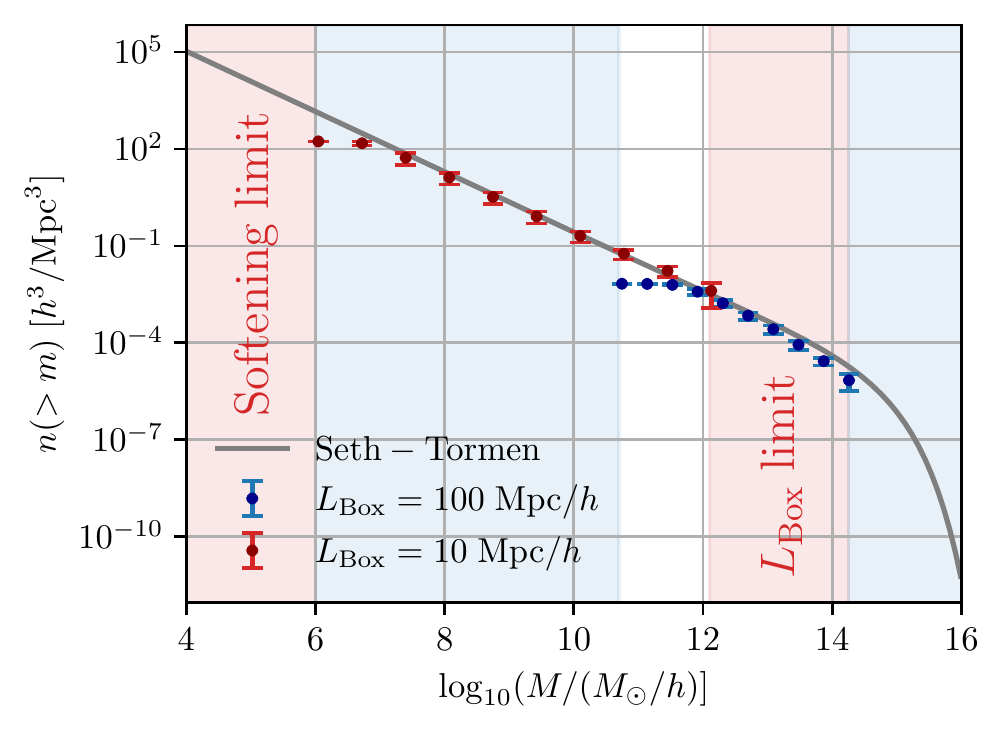}
    \caption{Halo mass function for $f(Q)$ gravitation with $L_\mathrm{Box}=10\mathrm{\;Mpc}/h$ and theoretical prediction for HMF by Seth-Tormen}
    \label{fig:88}
\end{figure}

Now we correspondingly derive the well-known halo mass function (further - just HMF), that obviously define the number of halos at a certain mass. Firstly, we built the halo catalogue for all of our snapshots with the use of halo/subhalo structure finder, namely \texttt{ROCKSTAR} (for more information on the subject, refer to the documentation paper \citep{Behroozi:2011ju}). Afterwards, we built the binned halo mass function, which is based on the values of $M_{\mathrm{200c}}$ (the mass of enclosed halo volume with energy density 200 times bigger that critical density of the universe $\rho_{\mathrm{cr}}$). Results are respectively plotted on the Figure (\ref{fig:88}) with the added Seth-Tormen theoretical prediction for halo mass function, based on \textit{Planck2018} fiducial cosmology and \texttt{CAMB} power spectrum at the $z=0$. Seth-Tormen HMF were computed using the python package \texttt{pynbody} \citep{2013ascl.soft05002P}.

From the Figure (\ref{fig:88}) shown above one could easily notice that generally our prediction for halo mass function from the MOG N-body simulation shows values of $n$ that approximately coincide with the ones that theoretically predicted by the Seth-Tormen HMF.
\section{LSS with large $L_{\mathrm{Box}}$: comparison}\label{sec:7}
\begin{figure*}
    \centering
    \includegraphics[width=\textwidth]{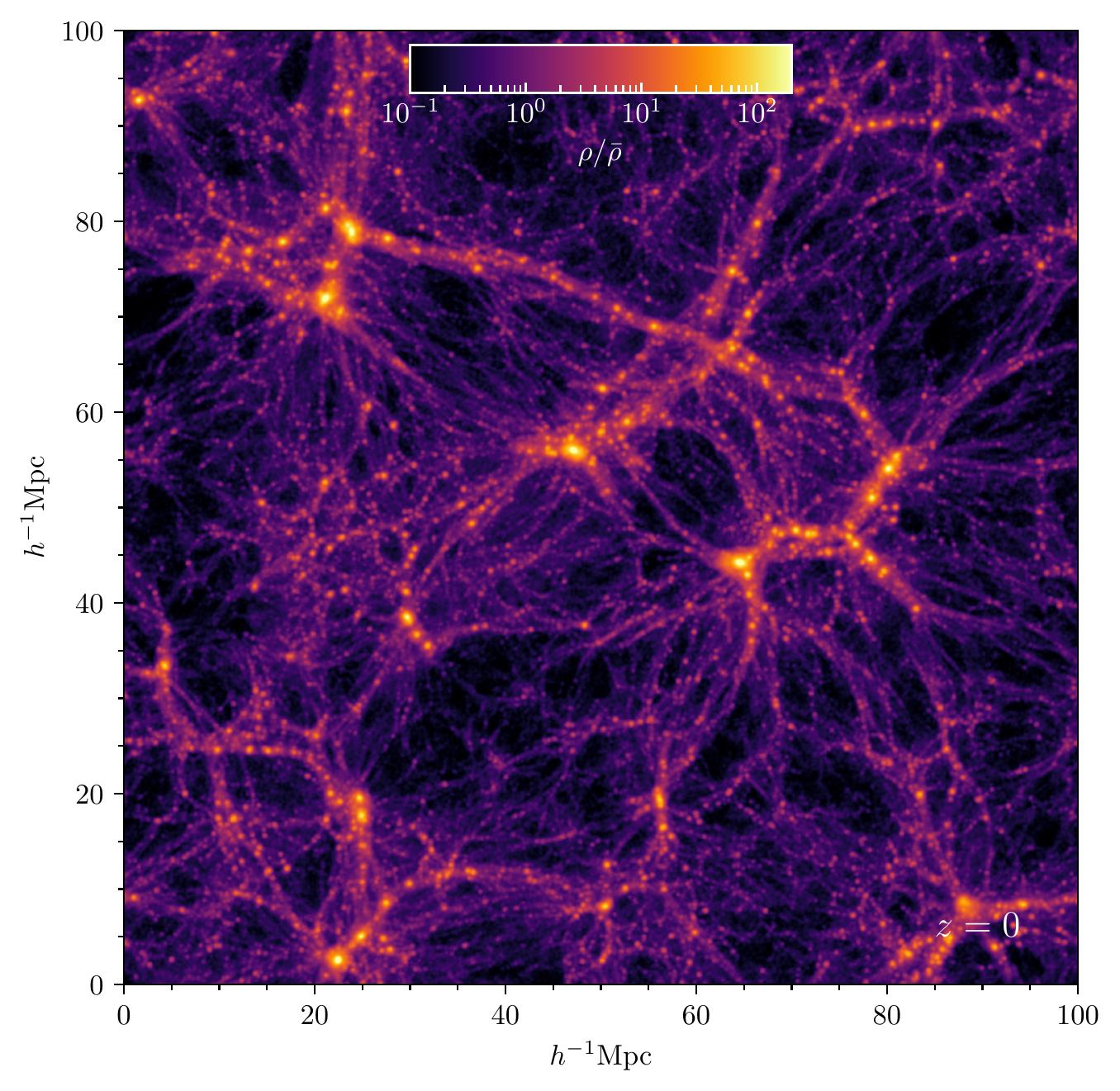}
    \caption{Present day snapshot of CDM over density for $L_{\mathrm{box}}=100h^{-1}$Mpc run}
    \label{fig:large}
\end{figure*}

As we previously mentioned, now we are going to perform an analysis of N-body simulation for bigger simulation box size, namely with $L_{\mathrm{box}}=100h^{-1}$Mpc. In that case, we only differ force resolution ($\epsilon=3.9$kpc), other cosmological parameters are assumed to be the same. At first, we as usual plot the CDM over density field for vanishing redshift at the Figure (\ref{fig:large}).

\begin{figure}
    \centering
    \includegraphics[width=\columnwidth]{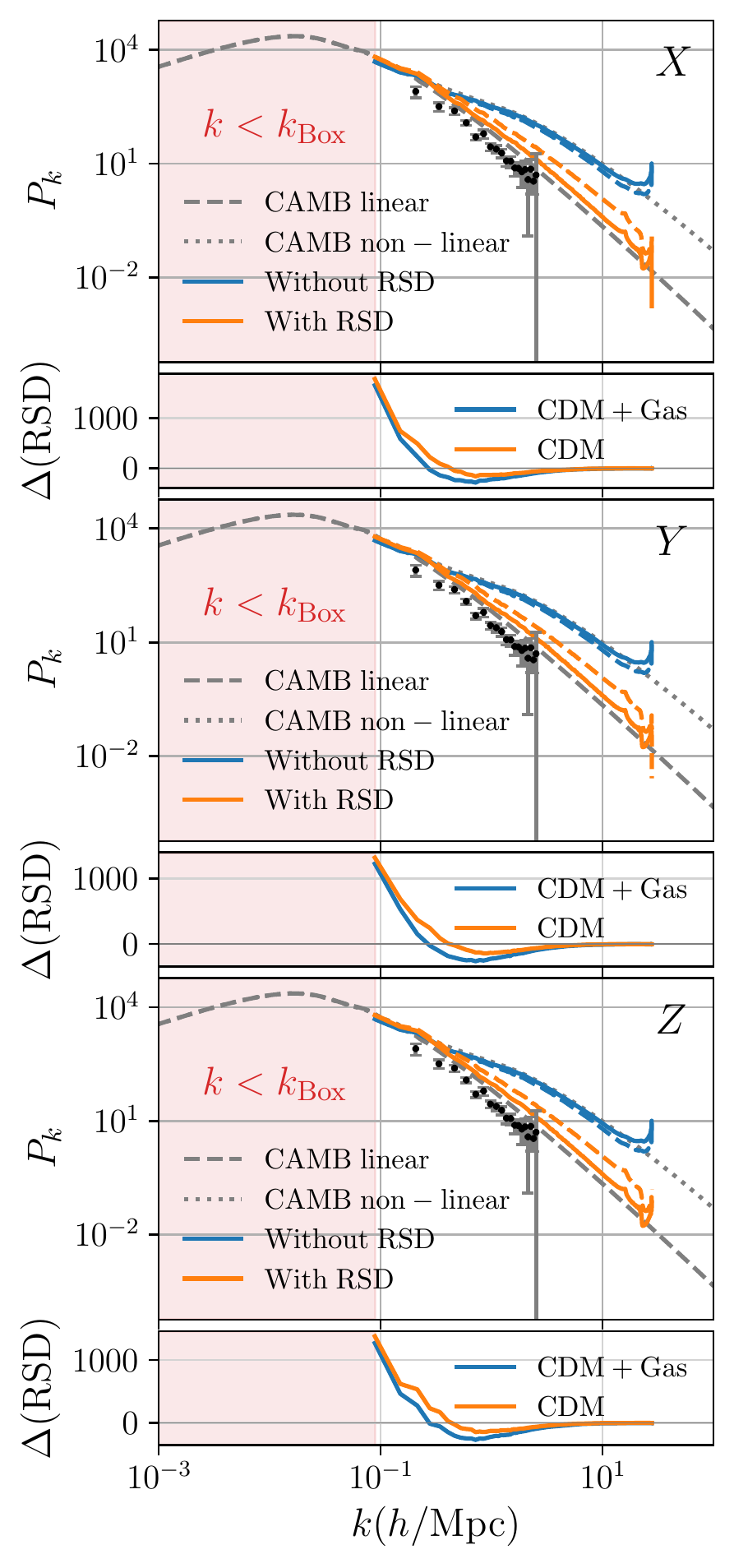}
    \caption{Matter power spectrum with/without RSDs for $f(Q)$ gravity vs. \texttt{CAMB} linear/non-linear $P(k)$ for $\Lambda$CDM. Dashed N-body $P(k)$ represent the CDM-only power spectrum, while solid line represent CDM+Gas $P(k)$. Error bars represent Ly $\alpha$ forest observations on high $z$. For this case, we have assumed large simulation box size of 100$h^{-1}$Mpc}
    \label{fig:Pk}
\end{figure}

In addition, we also plot the matter power spectrum for CDM, CDM+baryons on the Figure (\ref{fig:Pk}). As an obvious consequence of a larger box size, one can notice that maximum wave number $k$ grew to $k_{\mathrm{max}}\approx20\;h/$Mpc. Even at such big scales, our matter power spectrum, derived from the corresponding N-body simulation converge with theoretical prediction from \texttt{CAMB} code with up to sub-percent accuracy. As we noticed previously for small simulation box, axis of redshift-space distorsions had a vary small impact on the matter power spectrum. This statement holds for large $L_{\mathrm{box}}$ as well.

In the previous section we discussed the halo mass function for the case with smaller simulation box. Now we can discuss the same matter but for the larger $L_{\rm Box}$. As it appears, HMF extracted from the simulation replicates the Seth-Tormen HMF almost perfectly up to $M\approx 10^{14}M_{\odot}$ (see Figure (\ref{fig:88})). However, at bigger halo masses, our simulation HMF slightly differs from the theoretical prediction, which is usually observed in N-body simulations.
\begin{figure}
    \centering
    \includegraphics[width=\columnwidth]{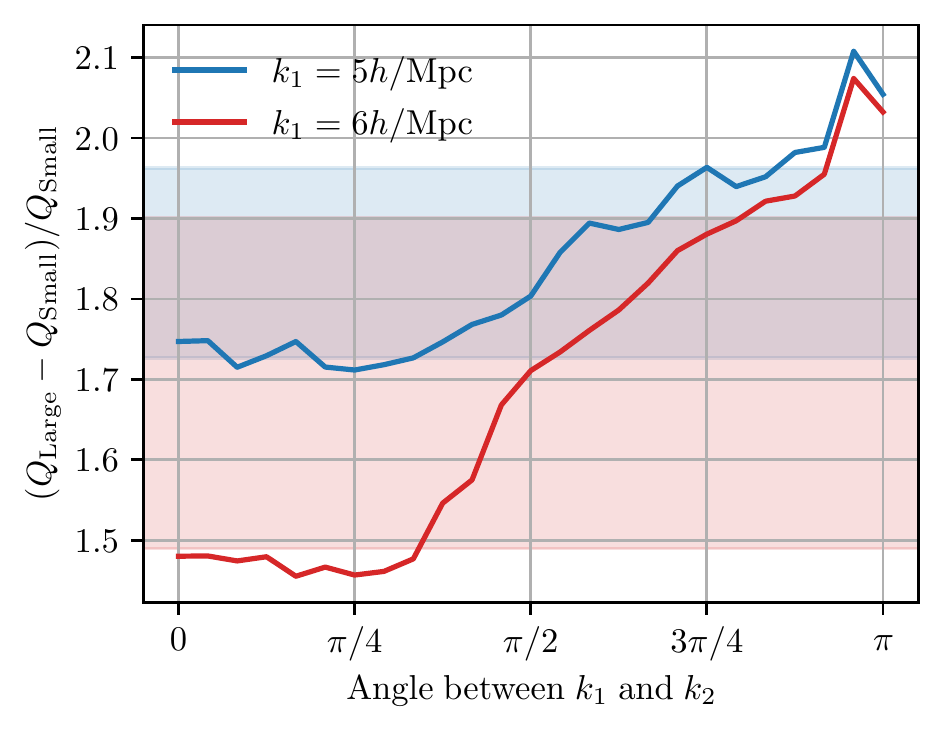}
    \caption{Relation of reduced matter power spectrum $Q(k_1,k_2,k_3)$ for both small and large simulation volumes with $1\sigma$ deviation, varying $k_1=2k_2$}
    \label{fig:Qk}
\end{figure}
\begin{figure*}
    \centering
    \includegraphics[width=\textwidth]{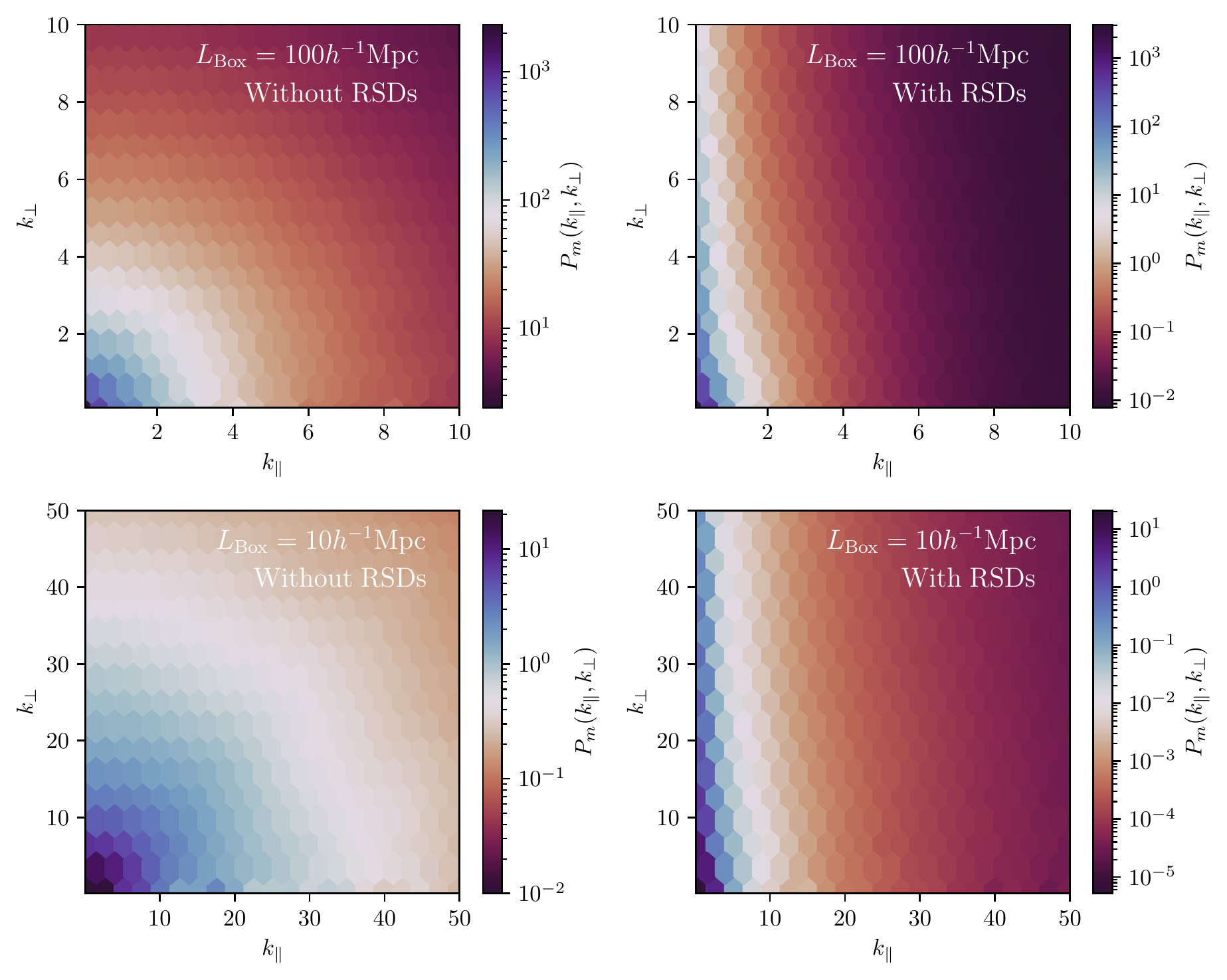}
    \caption{Two-dimensional matter power spectrum for small and large simulations within $f(Q)$ gravitation with/without RSDs}
    \label{fig:2D}
\end{figure*}
\subsection{Reduced bispectrum from 3PCF}
Finally, we also are going to introduce the so-called reduced bispectrum, which is derived from the regular bispectrum and matter power spectrum via the following relation, written down below:
\begin{equation}
    Q = \frac{B}{P_1P_2+P_2P_3+P_1P_3}
\end{equation}
Where $P_i=P_m(k_i)$. We consequently plot the relation between bispectrum of large and small cosmological volumes on the Figure (\ref{fig:Qk}). It is easy to notice that for smaller wave number ($k_1=5$), relation between $Q(k_1,k_2,k_3)$ for both cases has a mean value of $\approx1.9$ for all bins of angle $\theta$ (where it's maximum value is $\theta=\pi$, which is the angle between two sides of triangle $k_1$ and $k_2$). On the other hand, for relatively big $k_1$ (in our case, it's $k_1=6h/\rm Mpc$), deviation of large $L_{\mathrm{box}}$ reduced bispectrum from the small one is smaller because of the the fact that wave number span is shifted towards bigger values, while in the first case $k_1=5h/\rm Mpc$ were on the box size limit for smaller simulation, which distorted the results and caused the deviation to grow.

\subsection{2PCF for $f(Q)$ gravitation}
In addition to the matter power spectrum/reduced bispectrum and halo mass function, we as well derive the two point correlation function (further - just 2PCF) in a real space for CDM halos. Generally, 2PCF is defined as follows:
\begin{equation}
    \xi(|\mathbf{x}_1-\mathbf{x}_2|) = \langle \delta_m(\mathbf{x}_1)\delta_m(\mathbf{x}_2)\rangle
\end{equation}
Where $\mathbf{x}_i$ is three-dimensional position of an $i$-th CDM halo and $\delta_m$ is CDM over density parameter. We show the monopole redshift-space distorted two point correlation functions for both large and small simulations boxes on the Figure (\ref{fig:corrfunc}), where we added Quijote simulations \citep{Villaescusa_Navarro_2020} 2PCF, that admits \textit{Planck} fiducial cosmology. For the sake of completeness, we additionally marked BAO bumps for each case with color-coded dotted line. It is obvious, that in the case of small simulation box size, permitted range of $R$ is very small (up to $R\approx 2\times10^0h^{-1}$Mpc) and because of the small box size, correlation function is under sampled and does not coincide with Quijote one. On the other hand, for $L_{\rm Box}=100h^{-1}$Mpc simulation, correlation function corresponds to the Quijote one with sub-percent accuracy for range $R\in[2\times10^0,10^1]$. Now, we can proceed to the latest topic of our consideration, namely two-dimensional power spectra.
\subsection{2D matter power spectrum}
We plot the two-dimensional matter power spectrum for small and large boxes on the Figure (\ref{fig:2D}) with/without the presence of redshift-space distortions. As one can easily notice, on the plots with RSDs, the so-called "Finger of God" effect is observed, which arise because of the large scatter of galaxies recessional velocities at the small scales. Also, it is worth to inform that 2D matter power spectra for both box sizes are very alike. Now, since we discussed all of the topics for both simulation volumes within the modified theory of gravitation, we can proceed to the concluding remarks on the key findings within our study.

\section{Conclusions}\label{sec:8}
One can describe gravity using several geometric bases. The STGR, which attributes gravity to the nonmetricity tensor, has recently drawn much attention. A fascinating method for studying modified gravity is $f(Q)$ gravity, an extension of STGR. This study examined large scale structure formation observables using N-body simulations of $f(Q)$ gravitation for the first time to assess the theory's validity to cosmological context. Simulations were run with the use of \texttt{ME-GADGET} code, modification of the widely known \texttt{GADGET-2} code for two simulation boxes, namely $L_{\rm Box} = 10h^{-1}$Mpc and $L_{\rm Box} = 100h^{-1}$Mpc to decide on an optimal box size and compare the results for both simulation volumes.

We first performed Markov Chain Monte Carlo (MCMC) analysis for our exponential $f(Q)$ model to obtain best-fit values of MOG free parameters in Section (\ref{sec:4}). To test the fits provided by MCMC, we obtained theoretical predictions for the dimensionless mass densities $\Omega_{m0}$ and $\Omega_{\Lambda0}$, Hubble parameter $H(z)$, deceleration parameter $q(z)$ and statefinder pair $\{r,s\}$, $Om(z)$ parameter, placing graphical results on the Figures (\ref{fig:omega}), (\ref{fig:2}) and (\ref{fig:3}) respectively. As we noticed, Hubble parameter respected low redshift observations as well as deceleration parameter provided correct values of $q_0$ and transitional redshift within the constrained range. Moreover, statefinder diagnostics predict that initially universe was in Quintessence phase, passed the $\Lambda$CDM state up to Quintessence again. Finally, $Om(z)$ demonstrated that at the high-$z$ range, our universe was filled with phantom fluid, passed $\Lambda$CDM EoS at $z\approx2$ and now again has a phantom equation of state. After theoretical predictions, we started working on the N-body simulations whose primary findings corresponding to the quantities of interest are as follows:
\begin{itemize}
    \item Three-dimensional matter power spectrum monopole $P_k$: this was the first probe of a large-scale structure we studied in the present work. We plotted non-linear matter power spectra (with/without RSDs) for both small and large simulation volumes on the Figures (\ref{fig:77}) and (\ref{fig:Pk}) respectively, where we plotted CAMB linear/non-linear fiducial power spectra and observational data from Ly-$\alpha$ forest for the sake of comparison. One can notice that within the permitted range of wave number $k$ (limited by mean inter particle separation and simulation box length), non-linear matter power-spectra from small/large N-body simulations coincide with the CAMB one. However, for $L_{\rm Box}=10h^{-1}$Mpc, non-linear $P_k$ coincide with linear CAMB prediction too early because of the small box size.
    \item Halo Mass Function: the second significant quantity that can solely conclude the validity of a simulation. We place Seth-Tormen's theoretical HMF and the ones extracted from our N-body simulations in Figure (\ref{fig:88}). As we found, our small box size can cannot provide sufficient halo masses and reproduce viable halo mass function at all mass ranges up to the resolution limit, while large simulation follows Seth-Tormen prediction very precisely within the large span of halo masses $\log_{10}M/M_{\odot}\in[10,14]$, but gets slightly smaller than theoretical prediction for higher masses.
    \item Two-Point Correlation Function monopole $\xi_0(r)$: we as well investigate the redshift-space distorted correlation function monopoles in the Figure (\ref{fig:corrfunc}), where Quijote simulations correlation function is plotted to compare our results to fiducial cosmology. As remarked during numerical analysis, small box simulation fails to predict correct CDM halo correlations. On the other hand, in the range, $R\in[2\times10^0,10^1]h^{-1}$Mpc large box simulation precisely reconstructs Quijote data.
    \item Reduced bispectrum $Q(k_1,k_2,k_3)$: for the reduced bispectra case, we plotted the relation $(Q_{\rm Large}-Q_{\rm Small})/Q_{\rm Small}$ with different $k_1$ values (which acts as a triangle side length) on the Figure (\ref{fig:Qk}). We observed that this relation is generally close to $\approx 1.5$ across all bins of the angle between $k_1$ and $k_2$ (namely $\theta$) if one will assume value of $k_1$ that is not on the resolution limit for both cases (while it is worth to notice that we only adopted the case, where $k_2=2k_1$).
    \item Two-dimensional matter power spectrum $P_m(k_\parallel,k_\perp)$: This is the last quantity extracted from our N-body simulations. We plotted 2D power spectra for both sims in Figure (\ref{fig:2D}) with/without redshift-space distortions. From the plots, we noticed the so-called "Finger of God" effect that is present in the RSD case because of the elongated positions of CDM halos.
\end{itemize}
In conclusion, considering all the above points, the small simulation volume experiment failed to recreate matter power spectrum and correlation function correctly. However, second one, namely more extensive N-body simulation provided both viable 3D/2D matter power spectrum, correlation function, and halo mass functions and, therefore, we can consider exponential $f(Q)$ model to be a viable substitution of fiducial $\Lambda$CDM cosmology, since it not only satisfy many large scale structure constraints mentioned above, but also provide correct distance modulus up to high redshift values, where $\Lambda$CDM fails.

In the following papers, it will be interesting to investigate this modified gravity model using hybrid N-body and SPH simulations that incorporate supernova/AGN feedback, star and galaxy formation, jets, etc. using code \texttt{GIZMO} that allows the use of tabulated Hubble parameter and effective gravitational constant. It will, however, require a lot more computational resources (on the scale of millions of CPU hours).

\section*{Data Availability Statement}

There are no new data associated with this article.

\section*{Acknowledgements}
Sokoliuk O. performed the work in frame of the "Mathematical modeling in interdisciplinary research of processes and systems based on intelligent supercomputer, grid and cloud technologies" program of the NAS of Ukraine. The authors gratefully acknowledge the computing time provided on the high performance computing facility, Sharanga, at the Birla Institute of Technology and Science - Pilani, Hyderabad Campus. PKS  acknowledges the Science and Engineering Research Board, Department of Science and Technology, Government of India for financial support to carry out the Research project No.: CRG/2022/001847. SA acknowledges BITS-Pilani, Hyderabad campus for the Institute fellowship. We are very much grateful to the honorable referee and to the editor for the illuminating suggestions that have significantly improved our work in terms of research quality, and presentation. 



\bibliographystyle{mnras}
\bibliography{mybib} 





\bsp	
\label{lastpage}
\end{document}